\documentclass[%
 aip,
 amsmath,amssymb,
]{revtex4-1}

\usepackage{graphicx}
\usepackage{dcolumn}
\usepackage{bm}

\usepackage{amssymb}
\usepackage{amsmath}
\usepackage{color}
\usepackage{xcolor}
\usepackage[hidelinks]{hyperref}
\usepackage{multirow}
\usepackage{makecell}
\usepackage{booktabs}
\usepackage{lineno}

\usepackage[utf8]{inputenc}
\usepackage[T1]{fontenc}
\usepackage{mathptmx}
\usepackage{etoolbox}
\usepackage{natbib}

\makeatletter
\def\@email#1#2{%
 \endgroup
 \patchcmd{\titleblock@produce}
  {\frontmatter@RRAPformat}
  {\frontmatter@RRAPformat{\produce@RRAP{*#1\href{mailto:#2}{#2}}}\frontmatter@RRAPformat}
  {}{}
}%
\makeatother
\begin{document}

\preprint{AIP/123-QED}

\title{Finite difference form of the continuity condition at the polar axis, with application to the gyrokinetic simulation in a magnetic fusion torus}
\author{T. Wu}
\affiliation{Department of Engineering and Applied Physics, University of Science and Technology of China, Hefei, 230026, China}
\affiliation{National Key Laboratory of Frontier Physics in Controlled Nuclear Fusion, Chinese Academy of Sciences, Hefei, 230031, China}

\author{Z. Wang}%
\affiliation{Department of Engineering and Applied Physics, University of Science and Technology of China, Hefei, 230026, China}
\affiliation{National Key Laboratory of Frontier Physics in Controlled Nuclear Fusion, Chinese Academy of Sciences, Hefei, 230031, China}

\author{S. Wang\textsuperscript{*}}
\email{wangsj@ustc.edu.cn}
\affiliation{Department of Engineering and Applied Physics, University of Science and Technology of China, Hefei, 230026, China}
\affiliation{National Key Laboratory of Frontier Physics in Controlled Nuclear Fusion, Chinese Academy of Sciences, Hefei, 230031, China}

\date{\today}

\begin{abstract}
A new computational method to solve the hyperbolic (gyrokinetic Vlasov) equation and the elliptic (Poisson-like) equation at the polar axis is proposed. It is shown that the value of a scalar function at the polar axis can be predicted by its neighbouring values based on the continuity condition. This continuity condition systematically solves the pole problems including the singular factor $1/r$ in the  hyperbolic equation and the inner boundary in the elliptic equation. The proposed method is applied to the global gyrokinetic simulation of the tokamak plasma with the magnetic axis included.
\end{abstract}

\maketitle

\section{\label{sec1}Introduction}
The difficulties in numerically solving partial differential equations (PDEs) at the polar axis in polar coordinates have attracted significant interest for many years. These difficulties, which are noted as pole problems in this paper, are related to (i) terms containing the geometrical singular factor \cite{CONSTANTINESCU2002} $1/r$, with $r$ the radial position, (ii) inner boundary conditions \cite{GRIFFIN1979,HUANG1993} needed to be specified at $r=0$, even if physically there is no boundary at the polar axis. 

\begin{table} [htbp]
\resizebox{\textwidth}{!}{
\centering
    \begin{tabular}{|c|c|c|c|}
    \hline
    Framework & Reference & Problem & Note \\
    \hline
    \multirow{3}{*}{PSM} 
    & \multicolumn{1}{l|}{\multirow{1}{*}{Huang and Sloan (1993)}}             
    & \multicolumn{1}{l|}{\multirow{1}{*}{(i),(ii)}}
    & \multicolumn{1}{l|}{\multirow{1}{*}{Pole conditions }}  \\\cline{2-4}
    & \multicolumn{1}{l|}{\multirow{2}{*}{Matsushima and Marcu (1995)}}       
    & \multicolumn{1}{l|}{\multirow{2}{*}{(i),(ii)}}
    & \multicolumn{1}{l|}{\multirow{1}{*}{Spectral series that satisfy the continuity}} \\
    &                    
    &
    & \multicolumn{1}{l|}{\multirow{1}{*}{condition and have high convergence rate}} \\
    \hline
    \multirow{5}{*}{FDM} 
    & \multicolumn{1}{l|}{\multirow{2}{*}{Mohseni and Colonius (2000)}}   
    & \multicolumn{1}{l|}{\multirow{1}{*}{(i)}}
    & \multicolumn{1}{l|}{\multirow{1}{*}{Avoid setting grid points at the polar axis}} \\\cline{3-4}
    &                   
    & \multicolumn{1}{l|}{\multirow{1}{*}{(ii)}} 
    & \multicolumn{1}{l|}{\multirow{1}{*}{Computational domain mapping $\boldsymbol{M}$}} \\\cline{2-4}
    & \multicolumn{1}{l|}{\multirow{3}{*}{Constantinescu and Lele (2002)}}   
    & \multicolumn{1}{l|}{\multirow{2}{*}{(i)}}
    & \multicolumn{1}{l|}{\multirow{1}{*}{Using series expansions of physical quantities}} \\
    &                   
    &
    & \multicolumn{1}{l|}{\multirow{1}{*}{to analytically cancel singular factors in PDEs}} \\\cline{3-4}
    &                   
    & \multicolumn{1}{l|}{\multirow{1}{*}{(ii)}} 
    & \multicolumn{1}{l|}{\multirow{1}{*}{Computational domain mapping $\boldsymbol{M}$}} \\
    \hline
    \end{tabular}%
    }
\caption{\label{TablePreviousLit} Previous computational methods for solving pole problems in the frameworks of PSM and FDM. }
\end{table}

In the framework of the pseudo-spectrum method (PSM), several computational methods have been proposed to solve the pole problems, as shown in Table \ref{TablePreviousLit}. Under the assumption that the numerical solution is smooth near the polar axis in the polar coordinates, Huang and Sloan \cite{HUANG1993} constructed the pole condition to solve Problems (i) and (ii). Matsushima and Marcus \cite{MATSUSHIMA1995} solved Problems (i), (ii) by selecting the appropriate spectral series that satisfy the continuity condition and have high convergence rate. These spectral series analytically remove the singularity at the polar axis and ensure the high accuracy of the numerical solution near the polar axis.

Although the finite-difference method (FDM) is less accurate than the PSM, the treatment of pole problems in the FDM is of significant interest, due to its conveniences in handling complex geometrical configurations \cite{CONSTANTINESCU2002} and nonlinear computation. As shown in Table \ref{TablePreviousLit}, Mohseni and Colonius \cite{MOHSENI2000787} solved Problem (i) by avoiding  setting the grid points at the polar axis; the first point off the axis is situated at $\Delta r/2$, with $\Delta r$ the interval of the radial grid points. Constantinescu and Lele \cite{CONSTANTINESCU2002} solved Problem (i) by using the series expansions of physical quantities at the polar axis, which are derived from the continuity condition; these series expansions analytically cancel the singular factors in PDEs at the polar axis. For solving Problem (ii), the computational domain mapping $\boldsymbol{M}: (0,1)\times(-\pi,\pi) \to (-1,1)\times(\pi/2,\pi/2)$ was used, which avoids the inner boundary condition by using the one-sided finite-difference method \cite{GRIFFIN1979}.

\begin{figure}
\centering
\includegraphics[width=0.8\linewidth]{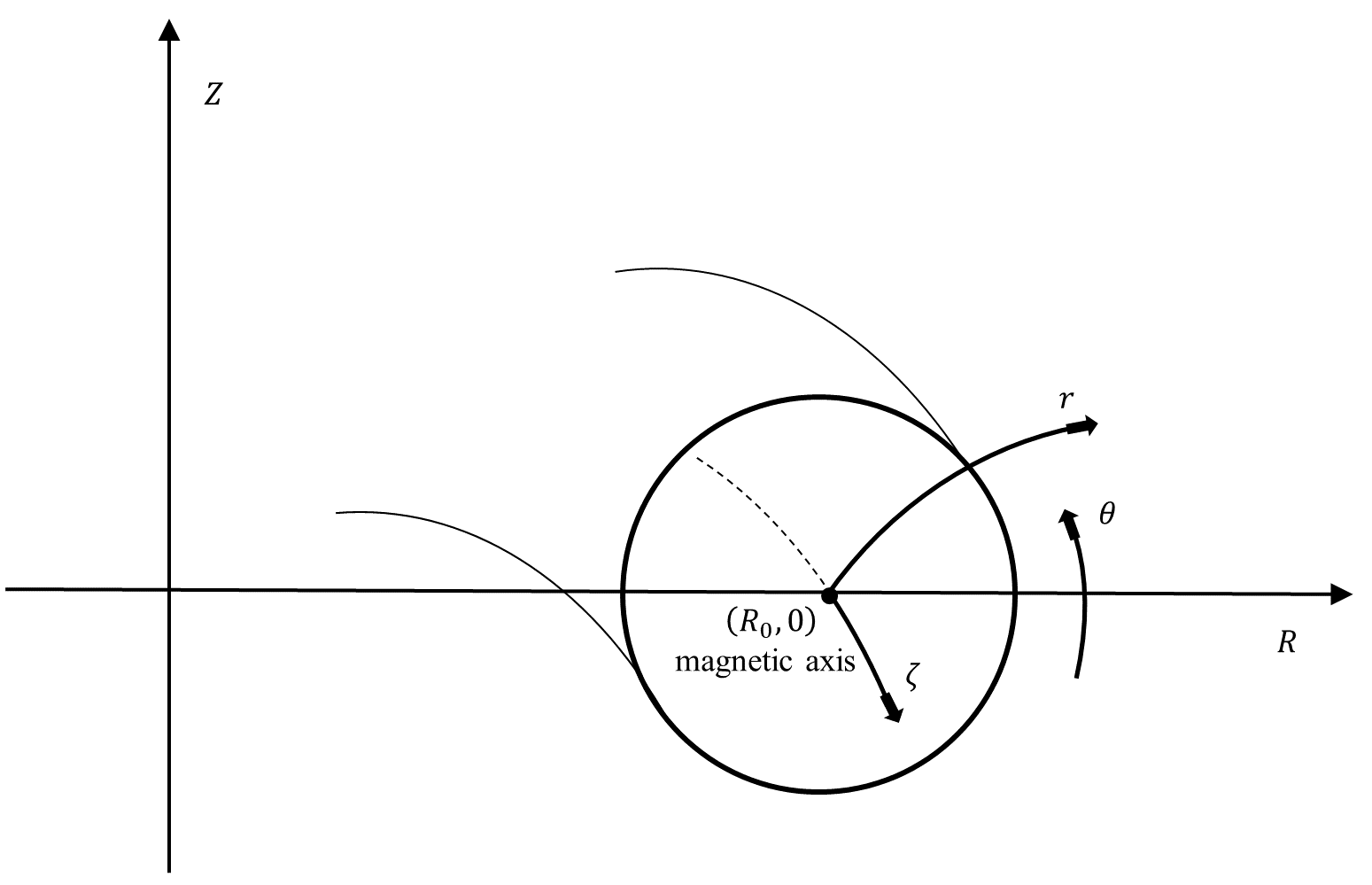}
\caption{\label{FigTokamak} Magnetic flux coordinates and cylindrical coordinates.  }
\end{figure}

The pole problems are also of interest in the global gyrokinetic (GK) simulation \cite{McClenaghan2014POP,Bouzat2018ESAIM,WANG2023} in the tokamak fusion plasma, since the magnetic axis of the fusion torus is essentially a polar axis. To simulate the ion temperature gradient (ITG) driven mode in a fusion torus with adiabatic electrons, one has to solve the GK Vlasov-Poisson (VP) system. However, the GK VP system is very computationally expensive, which simulates the time evolution of the distribution function $F(\boldsymbol{Z};t)$ and the perturbed electrostatic potential $\delta \phi(\boldsymbol{X};t)$, where $\boldsymbol{Z}=(\boldsymbol{X},v_{\parallel},\mu)$ are the phase space coordinates, $\boldsymbol{X}$ is the position of the gyrocenter, $v_{\parallel}$ is the the parallel velocity and $\mu$ is the magnetic moment with $\mathrm{d}\mu/\mathrm{d}t=0$. The system is usually solved in magnetic flux coordinates $\boldsymbol{X}=(r,\theta,\zeta)$, with $r(\psi_{T})\propto\sqrt{\psi_{T}}$ the generalized minor radius, $\psi_{T}$ the toroidal magnetic flux, $\theta$ the poloidal angle and $\zeta$ the toroidal angle. The magnetic flux coordinates discussed are graphically shown in Fig. \ref{FigTokamak}. In these coordinates, the GK Vlasov equation \cite{BRIZARD2007} is given by

\begin{align} \label{EqVlasov}
    \frac{\partial F}{\partial t}+ \Dot{r} \frac{\partial F}{\partial r} + \Dot{\theta} \frac{\partial F}{\partial \theta} + \Dot{\zeta} \frac{\partial F}{\partial \zeta} + \Dot{v}_{\parallel} \frac{\partial F}{\partial v_{\parallel}} = 0,
\end{align}
which is a hyperbolic equation. The GK quasi-neutrality equation \cite{LEE1983} is given by

\begin{align} \label{EqQn}
    -c_{1}\delta \phi +  \int{\mathrm{d}^{3}v \left( \frac{c_{1}}{n_0}F \right) \left<\left< \delta \phi \right>\right>_{ga}} - c_{2}(\delta \phi-\left< \delta \phi \right>_{FA}) = -e_{i} \rho_{i,gy},
\end{align}
with $c_{1}=\frac{e_{i}^2n_{0}}{T_{i}}$, $c_{2}=\frac{e^2n_{0}}{T_{e}}$, $\left< \cdot \right>_{FA}$ the magnetic surface averaged operator and $\left< \cdot \right>_{ga}$ the gyro-average operator. Here $n_{0}$ is both the equilibrium ion density and the equilibrium electron density, $m_i$ is the ion mass. $T_{i}$ and $e_{i}$ are the ion temperature and charge, respectively; $T_{e}$ and $e$ are the electron temperature and charge, respectively. In the long-wavelength approximation, the GK quasi-neutrality equation becomes the GK Poisson (elliptic) equation \cite{LEE1983}

\begin{align} \label{EqPoisson}
    \nabla \cdot \left( c_{0} \nabla_{\perp} \delta \phi  \right) - c_{2}(\delta \phi-\left< \delta \phi \right>_{FA}) = -e_{i} \rho_{i,gy},
\end{align}
with $c_{0}=\frac{n_{0}m_i}{B^{2}_{0}}$ and $B_{0}$ the magnetic field at the magnetic axis. When approaching the magnetic axis, $r\to0$, and $r$ can be understood as the usual minor radius. On a minor cross section, the transformation from magnetic flux coordinates $(r,\theta)$ to cylindrical coordinates $(R,Z)$ near the magnetic axis is given by

\begin{align} \label{EqMapping}
    R-R_{0} = x = r \cos{\theta},  \\
    Z = y = r \sin{\theta},  \notag
\end{align}
where $(R-R_{0},Z)$ or $(x,y)$ are the pseudo-Cartesian coordinates \cite{Jolliet2009,Lapillon2010,Lanti2019}, and $(r,\theta)$ are essentially the polar coordinates. So the difficulties in the GK simulation at the magnetic axis are essentially the pole problems at the polar axis. When Eq. \eqref{EqPoisson} is numerically solved with $r\to 0$, $\Dot{\theta}\to \infty$ is a singular term. This is Problem (i) in the GK Vlasov (hyperbolic) equation. The inner boundary at the magnetic axis in Eq. \eqref{EqPoisson} is Problem (ii) in the Poisson (elliptic) equation. To include the magnetic axis in the GK simulation, we are forced to solve these problems. Note that previous global GK simulations \cite{LIN1998,CANDY2003,Grandgirard2006} excluded the magnetic axis from the simulation domain, and an inner boundary condition is needed.

A number of GK codes, such as ORB5 \cite{JOLLIET2007}, GT5D \cite{IDOMURA2008}, GTC \cite{McClenaghan2014POP}, GKNET \cite{OBREJAN2015}, GYSELA \cite{Bouzat2018ESAIM} and NLT \cite{DAI2019}, have been updated to include the magnetic axis. The computational method used in these codes are shown in Table \ref{TablePreviousGKcodes}, and are briefly summarized as follows. 

\begin{table} [htbp]
\resizebox{\textwidth}{!}{
\centering
    \begin{tabular}{|c|c|c|c|}
    \hline
    Code & Reference & Problem & Note \\
    \hline
    \multirow{2}{*}{ORB5}     
    & \multicolumn{1}{l|}{\multirow{2}{*}{Jolliet \textit{et al.} (2007)}}             
    & \multicolumn{1}{l|}{\multirow{1}{*}{(i)}}
    & \multicolumn{1}{l|}{\multirow{1}{*}{Lagrangian-PIC method, using the pesudo-Cartesian coordinates}}  \\\cline{3-4}
    &                  
    & \multicolumn{1}{l|}{\multirow{1}{*}{(ii)}} 
    & \multicolumn{1}{l|}{\multirow{1}{*}{Finite element method, using the regularity condition}} \\
    \hline
    \multirow{3}{*}{GT5D}     
    & \multicolumn{1}{l|}{\multirow{2}{*}{Idomura \textit{et al.} (2008)}}             
    & \multicolumn{1}{l|}{\multirow{2}{*}{(i)}}
    & \multicolumn{1}{l|}{\multirow{1}{*}{Eulerian method (finite difference method),}}  \\
    & \multicolumn{1}{l|}{\multirow{2}{*}{Matsuoka \textit{et al.} (2018)}}         
    &
    &\multicolumn{1}{l|}{\multirow{1}{*}{avoiding setting grid points at the magnetic axis}}  \\\cline{3-4}
    &
    & \multicolumn{1}{l|}{\multirow{1}{*}{(ii)}} 
    & \multicolumn{1}{l|}{\multirow{1}{*}{Finite element method, using the regularity condition}} \\
    \hline
    \multirow{3}{*}{GTC}     
    & \multicolumn{1}{l|}{\multirow{2}{*}{McClenaghan \textit{et al.} (2014)}}             
    & \multicolumn{1}{l|}{\multirow{1}{*}{(i)}}
    & \multicolumn{1}{l|}{\multirow{1}{*}{Lagrangian-PIC method, using the pesudo-Cartesian coordinates}}  \\\cline{3-4}
    & \multicolumn{1}{l|}{\multirow{2}{*}{Feng \textit{et al.} (2018)}}                  
    & \multicolumn{1}{l|}{\multirow{2}{*}{(ii)}} 
    & \multicolumn{1}{l|}{\multirow{1}{*}{FDM, using the linear boundary condition;}} \\
    &
    &
    & \multicolumn{1}{l|}{\multirow{1}{*}{finite element method, using the zero boundary condition}} \\
    \hline
    \multirow{2}{*}{GKNEK}     
    & \multicolumn{1}{l|}{\multirow{2}{*}{Obrejan \textit{et al.} (2015)}}             
    & \multicolumn{1}{l|}{\multirow{2}{*}{(i)}}
    & \multicolumn{1}{l|}{\multirow{1}{*}{Eulerian method  (finite difference method),}}  \\
    &
    &
    & \multicolumn{1}{l|}{\multirow{1}{*}{using the cylindrical coordinates}} \\
    \hline
    \multirow{2}{*}{GYSELA}     
    & \multicolumn{1}{l|}{\multirow{2}{*}{Bouzat \textit{et al.} (2018)}}             
    & \multicolumn{1}{l|}{\multirow{1}{*}{(i)}}
    & \multicolumn{1}{l|}{\multirow{1}{*}{Semi-Lagrangian method, using the pesudo-Cartesian coordinates}}  \\\cline{3-4}
    &                   
    & \multicolumn{1}{l|}{\multirow{1}{*}{(ii)}} 
    & \multicolumn{1}{l|}{\multirow{1}{*}{FDM, using the regularity condition}} \\
    \hline
    \multirow{2}{*}{NLT (previous)}     
    & \multicolumn{1}{l|}{\multirow{2}{*}{Dai \textit{et al.} (2019)}}             
    & \multicolumn{1}{l|}{\multirow{1}{*}{(i)}}
    & \multicolumn{1}{l|}{\multirow{1}{*}{FDM, using the cylindrical coordinates}}  \\\cline{3-4}
    &                   
    & \multicolumn{1}{l|}{\multirow{1}{*}{(ii)}} 
    & \multicolumn{1}{l|}{\multirow{1}{*}{FDM, using Gauss's theorem}} \\
    \hline
    \end{tabular}%
    }
\caption{\label{TablePreviousGKcodes} Previous computational methods for solving pole problems in the global GK codes including the magnetic axis. }
\end{table}

To solve Problem (i) in sovling the GK Vlasov equation,  two types of method are used. Type 1, the cylindrical coordinates or pseudo-Cartesian coordinates are used in the entire computational domain \cite{IDOMURA2008,OBREJAN2015,Bouzat2018ESAIM}, or are just used near the magnetic axis \cite{Jolliet2009,McClenaghan2014POP,Lanti2019,DAI2019}. Type 2, avoid setting the grid points at the magnetic axis \cite{Matsuoka2018POP}; the first grid point off the axis is situated at $\Delta r/2$. Note that the Type 1 method do not use the magnetic coordinates in the entire computational domain, however, the use of magnetic coordinates has some advantages \cite{JOLLIET2007}.

To solve Problem (ii) in sovling the GK Poisson equation, different kinds of inner boundary conditions at the magnetic axis have been applied. Regularity conditions, $\delta \phi(r=0,\theta,\zeta)=\delta \phi(r=0,0,\zeta)$, $\delta \phi(-\Delta r/2,\theta,\zeta)=\delta \phi(\Delta r/2,\theta+\pi,\zeta)$ and $\delta \phi((\frac{1}{2}-i)\Delta r,\theta,\zeta)=\delta \phi((i-\frac{1}{2})\Delta r,\theta+\pi,\zeta)$, are respectively used in ORB5 \cite{JOLLIET2007}, GYSELA \cite{Bouzat2018ESAIM} and GT5D \cite{Matsuoka2018POP}; the linear boundary condition \cite{McClenaghan2014POP} or the zero boundary condition \cite{FENG2018} is used in GTC. In the previous version of NLT \cite{DAI2019}, Problem (ii) is solved by using the Gauss's theorem, which avoids the use of inner boundary condition. 

Note that previously, the global GK codes solve Problem (i) and Problem (ii) by using different method; Problem (i) in the GK Poisson equation was usually solved in the cylindrical coordinates, however Problem (ii) in the GK Poisson equation was solved by using the magnetic (polar) coordinates with inner boundary conditions used; a coordinate transformation between cylindrical coordinates and magnetic surface coordinates is required to use these methods. In GT5D \cite{Matsuoka2018POP}, Problem (i) was solved in the magnetic coordinates. However, the first grid point off the axis is situated at $\Delta r/2$ as opposed to $\Delta r$, which may lead to more severe numerical errors or instabilities near the axis \cite{CONSTANTINESCU2002}.  

In this paper, a new computational method based on the FDM to solve the pole problems is proposed. It is found that the value of a scalar function at the polar axis can be predicted by its neighbouring values based on the continuity condition. Problem (i) and Problem (ii) are systematically solved in the polar coordinates by using this continuity condition. 

The proposed method is used to update the NLT code \cite{Ye2016,XU2017} to include the magnetic axis, which evolves the perturbed distribution function $\delta f$ along the equilibrium orbit by using the characteristic line method and takes account of the perturbation effects by using the numerical Lie transform \cite{WANG2012,WANG2013POP,Wang2013PRE,Xu2014POP}. For other global GK codes based on the FDM, such as Eulerian codes, the proposed method can easily be used to update these codes to include the magnetic axis.

The remaining part of this paper is organized as follows. In Section \ref{sec2}, the finite difference form of the continuity condition at the polar axis is presented. In Section \ref{sec3}, the application of the proposed method in GK simulation for a tokamak torus is presented. In Section \ref{sec4}, numerical results near the magnetic axis are presented. Finally, the conclusion is presented in Section \ref{sec5}. In addition to Problem (i) and Problem (ii), there may be Problem (iii): the severe numerical error associated with the low-order finite-difference schemes. We generalize the proposed method to mitigate the numerical error. The details of the generalized method is presented in Appendix \ref{SecGmeanValue}.

\section{Finite difference form of the continuity condition at the polar axis} \label{sec2}
In this section, the continuity condition at the origin in the Cartesian coordinates and in the polar coordinates is discussed, and its discretization form at the polar axis is presented.

\subsection{Continuity condition at the origin in the Cartesian coordinates and in the polar coordinates}
The functions to be solved in the hyperbolic equation and the elliptic equation discussed here are physically observable scalars. These equations in mathematical physics can be written in any coordinates. To proceed our discussion, we introduce a fundamental assumption 

\emph{Any physical observable scalar $g(\boldsymbol{P})$ is continuous in the Euclidean space, with $\boldsymbol{P}$ the space point.} 

Note that it is independent of the coordinate. This fundamental assumption shall be referred to as the "continuity condition".

The fundamental assumption implies that the function $g(x,y)$, which is of interest, is $\mathcal{C}^{\infty}$ at the origin of the Cartesian coordinates. So it can be Taylor expanded, to any desired accuracy, around the origin

 \begin{align} \label{EqCarSeris}
     g(x,y) = \sum_{j=0}^{\infty} \sum_{k=0}^{\infty} \frac{1}{j!k!} 
     a_{j,k}x^{j}y^{k},
 \end{align}
with $a_{j,k}=\frac{\partial^{j+k}g} {\partial^jx\partial^ky}$. \par
In different coordinates, the value of a scalar quantity at the same space point should be invariant. Transforming from the Cartesian to the polar coordinates (by using Eq. \eqref{EqMapping}), one finds that Eq. \eqref{EqCarSeris} can be written as 

 \begin{align} \label{EqPolSeries}
     \hat{g}(r,\theta) = \sum_{m=-\infty}^{\infty} e^{im\theta} r^{\lvert m \rvert}  \sum_{l=0}^{\infty}
     A_{m}^{(l)} r^{2l}  ,
 \end{align}
where we have regrouped terms with the same poloidal Fourier number $m$ together. Here, $A_{m}^{(l)}$ are the coefficients of the series expansion for $\hat{g}(r,\theta)$, which is expanded as a power series in $r$ and a Fourier series in $\theta$. The expression of the coefficients $A_{m}^{(l)}$ in terms of $a_{j,k}$ is not shown here, because it will not be used in our computational method. \par
The series expansion shown in Eq. \eqref{EqPolSeries} has been obtained in the previous literature \cite{LEWIS1990,EISEN1991}. Lewis \cite{LEWIS1990} derived Eq. \eqref{EqPolSeries} from the symmetry constraint in polar coordinates and the regularity constraint in Cartesian coordinates. Eisen et al. \cite{EISEN1991} proved that Eq. \eqref{EqPolSeries} could be derived from the regularity condition in the Cartesian coordinates. \par
We point out that Eq. \eqref{EqCarSeris} is the representation of the fundamental assumption of continuity condition in Cartesian coordinates, while Eq. \eqref{EqPolSeries} is its representation in the polar coordinates.

Eq. \eqref{EqPolSeries} can be written as $     \hat{g}(r,\theta) = \sum_{m=-\infty}^{\infty} \hat{g}_m(r)e^{im\theta}$, with
  \begin{align} \label{EqSingleMSeries}
      \hat{g}_{m}(r) = r^{\lvert m \rvert} \sum_{l=0}^{\infty}A_{m}^{(l)}r^{2l}.
  \end{align}
  It is clearly seen from Eq. \eqref{EqSingleMSeries} that powers of $r$ in $\hat{g}_{m}(r)$ are not smaller than $\lvert m \rvert$, and that $\hat{g}_{m}(r)$ is an even (odd) function when $m$ is an even (odd) number, which is the symmetry condition \cite{LEWIS1990}. 

\subsection{Discretization form at the polar axis}
  The continuity condition can be used to construct the numerical solution at the polar axis without solving the PDEs directly. According to Eq. \eqref{EqSingleMSeries}, only the $m=0$ component is nonzero at the polar axis, i.e., $\hat{g}(0,\theta)=\hat{g}_{0}(0)$. 
  $\hat{g}_{0}(0)$ can be predicted by the continuity condition. In the neighbourhood of the polar axis, one finds from Eq. \eqref{EqSingleMSeries} that   

  \begin{align} \label{EqTruncSeries}
      \hat{g}_{0}(r) = A_{0}^{(0)} + A_{0}^{(1)}r^2 + \cdots.
  \end{align}
The solution of $\hat{g}(0)$ requires the value of $A^{(0)}_{0}$, which can be calculated from $\hat{g}_0(\Delta r)$ and $\hat{g}_0(2\Delta r)$, with $\Delta r$ the interval of radial grid points. By truncating higher-order terms ($r^{l}$, with $l>2$)  in Eq. \eqref{EqTruncSeries}, one obtains

\begin{align}  \label{EqSolveCoe}
\begin{cases}
    A_{0}^{(0)} + A_{0}^{(1)} \Delta r^2 = \hat{g}_{0}(\Delta r) , \\
    A_{0}^{(0)} + 4A_{0}^{(1)} \Delta r^2 = \hat{g}_{0}(2\Delta r) .
\end{cases}
\end{align}
The truncation error is consistent with the error of the second-order central difference. Solving Eq. \eqref{EqSolveCoe} gives

  \begin{align} \label{EqM0Axis}
      \hat{g}_{0}(0) = A_{0}^{(0)} = \frac{4}{3}\hat{g}_{0}(\Delta r) - \frac{1}{3}\hat{g}_{0}(2\Delta r).
  \end{align}
The above equation suggests that $\hat{g}_{0}(0)$ can be predicted by given $\hat{g}_0(\Delta r)$ and $\hat{g}_0(2\Delta r)$. Writing Eq. \eqref{EqSolveCoe} in the dicretization form, one obtains

  \begin{align} \label{EqAxisValue}
      \hat{g}(0,\theta) = \frac{4}{3N_{\theta}}\sum_{k=1}^{N_{\theta}}\hat{g}(\Delta r,\theta_{k}) - \frac{1}{3N_{\theta}}\sum_{k=1}^{N_{\theta}}\hat{g}(2\Delta r,\theta_{k}),
  \end{align}
with $\theta_{k} = -\pi+(k-1)\Delta \theta$, $\Delta\theta=2\pi/N_{\theta}$ and $N_{\theta}$ the number of $\theta$ grid points. The uniform radial grid points are defined as $r_{j}=(j-1)\Delta r$, with $\Delta r = r_{b}/(N_{r}-1)$, $r_{b}$ the outer radial boundary and $N_{r}$ the number of radial grid points.

We note that Eq. \eqref{EqAxisValue} is the dicretization form of the continuity condition at the polar axis. 

\begin{figure}
\centering
\includegraphics[width=0.5\linewidth]{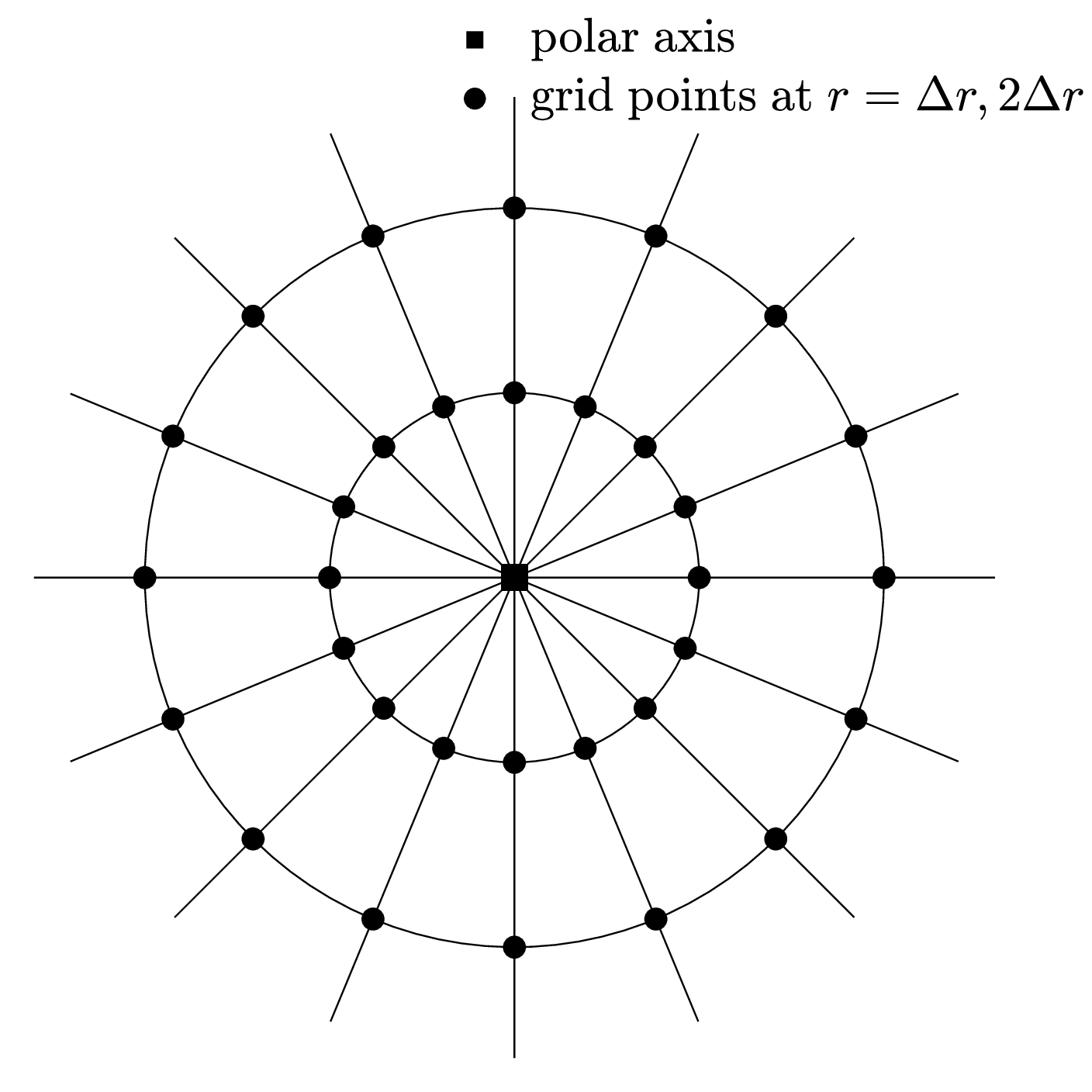}
\caption{\label{FigNumPoints} Predicting a scalar function at the polar axis by using the function's  average value on the neighbouring points.}
\end{figure}

Fig. \eqref{FigNumPoints} shows the grid points used in Eq. \eqref{EqAxisValue}, which indicates that the value of a scalar function at the polar axis can be predicted by its average value in the neighbouring area. Therefore, Eq. \eqref{EqAxisValue} can be referred to as the "mean value theorem". 

In solving the elliptic (Poisson-like) equation, Eq. \eqref{EqAxisValue} serves as the numerical inner boundary condition. This inner boundary condition is just derived from the continuity condition without any other additional assumptions (Problem (ii)); we note that the pole is not a boundary from the viewpoint of geometry or physics. Note that different inner boundary conditions were used in the global GK simulations. The regularity conditions \cite{JOLLIET2007,Matsuoka2018POP,Bouzat2018ESAIM} satisfy the continuity condition. However, the linear \cite{McClenaghan2014POP} and zero \cite{FENG2018} boundary conditions satisfy the continuity condition only for $m=1$ and $m\neq 0$, respectively.

In solving the hyperbolic (GK Vlasov) equation, the scalar function to be solved for at the polar axis can be predicted by the mean value theorem, without solving the hyperbolic equation itself directly at the pole; this method avoids the numerical treatment of the $1/r$ singularity term (Problem (i)).

\section{Application in the gyrokientic simulation for a tokamak torus} \label{sec3}

\subsection{Brief review of the NLT code} \label{SecNLT}
In the global GK code NLT \cite{Ye2016,XU2017}, the adiabatic electron assumption is used. The ion gyrocenter distribution function $F(\boldsymbol{Z},t)$ is divided into $F=F_{0}+\delta f$, with $\boldsymbol{Z}=(\boldsymbol{X},v_{\parallel},\mu)$ the five-dimensional gyrocenter phase space coordinates, $\boldsymbol{X}$ the three-dimensional space coordinates, and $F_{0}$/$\delta f$ the equilibrium/perturbed ion distribution function. The gyrocenter motion is divided into the perturbed $\boldsymbol{\dot{Z}_{1}}=\{\boldsymbol{Z},\delta h\}$ and the unperturbed one $\boldsymbol{\dot{Z}_{0}}=\{\boldsymbol{Z},H_0\}$, with $\{ \cdot\}$ the Poisson bracket, $\delta h/H_{0}$ the perturbed/unperturbed Hamiltonian. It is noted that $F$ is independent of the gyroangle $\xi$. The nonlinear GK Vlasov equation is written as

\begin{align} \label{EqFirstOdVlasov}
    \partial_{t}\delta f + \dot{\boldsymbol{Z}_{0}}\cdot\partial_{\boldsymbol{Z}} \delta f = -\dot{\boldsymbol{Z}_{1}}\cdot\partial_{\boldsymbol{Z}} F_{0}-\dot{\boldsymbol{Z}_{1}}\cdot\partial_{\boldsymbol{Z}} \delta f.
\end{align} 

The NLT method \cite{WANG2012,WANG2013POP,Wang2013PRE} decouples the perturbed gyrocenter
motion from the unperturbed one within a short time interval $\Delta t$, the time step in numerical computation, by using the numerical Lie-Transform method to remove the perturbed gyrocenter Hamiltonian. The six-dimensonal gyrocenter phase space coordinates $(\boldsymbol{Z},\xi)$ are transformed to the new coordinates $(\bar{\boldsymbol{Z}},\bar{\xi})$, with $\boldsymbol{\bar{Z}}=(\boldsymbol{\bar{X}},\bar{v}_{\parallel},\bar{\mu})$. In the new coordinates, the equation of motion is identical to the unperturbed one, which reads

\begin{align} \label{EqMotionofZbar}
    \boldsymbol{\dot{\bar{Z}}} = \boldsymbol{\dot{\bar{Z}}}_{0} = \{\bar{\boldsymbol{Z}},H_{0}(\bar{\boldsymbol{Z}})\},
\end{align}
with $H_{0}(\bar{\boldsymbol{Z}})=\frac{1}{2}m_{i}\bar{v}_{\parallel}^{2}+\bar{\mu}B$. $\dot{\bar{\xi}}$ is not discussed here, because $\boldsymbol{\dot{\bar{Z}}}$ are decoupled from $\bar{\xi}$. The coordinate transformation is given by \cite{WANG2013POP}

\begin{align} \label{EqLieTrans}
    \boldsymbol{\bar{Z}}=\boldsymbol{Z}+\boldsymbol{G}_{1}+\frac{1}2{}\boldsymbol{G}_{1}\cdot \nabla \boldsymbol{G}_{1},
\end{align}
where $\boldsymbol{G}_{1}$ is the first-order generating vector and is calculated from the first-order gauge function of the Lie-transform

\begin{align} \label{EqG1}
    \boldsymbol{G}_{1} = -\{\bar{\boldsymbol{Z}},S_{1}\}.
\end{align}
The gauge function $S_1$ is obtained by 

\begin{align} \label{EqS1}
    \frac{\mathrm{d_0}}{\mathrm{d}t}S_1(\bar{\boldsymbol{Z}},t) = \delta h(\bar{\boldsymbol{Z}},t),
\end{align}
with $\frac{\mathrm{d_0}}{\mathrm{d}t}=\partial_t+\dot{\bar{\boldsymbol{Z}}}_0\cdot\partial_{\bar{\boldsymbol{Z}}}$.

In the new coordinates $\bar{\boldsymbol{Z}}$, the GK Vlasov equation is given by 

\begin{align} \label{EqunPerturbedOrb}
    \frac{\mathrm{d_0}}{\mathrm{d}t}\delta \bar{f}(\bar{\boldsymbol{Z}},t) = 0.
\end{align}
It is noted that $\delta \bar{f}$ is independent of $\bar{\xi}$. It can be seen from Eq. \eqref{EqunPerturbedOrb} that $\delta \bar{f}$ evolves along the unperturbed orbit, which is solved by using the characteristic line method. In each given short time interval $[t_{i},t_{i+1}]$, the value of $\delta \bar{f}(\bar{\boldsymbol{Z}},t_{i+1})$ at the fixed phase space grid is obtained by retracing the fixed phase space point along the unperturbed orbit, which reads

\begin{align}
    \delta \bar{f}(\bar{\boldsymbol{Z}},t_{i+1}) = \delta \bar{f}(\bar{\boldsymbol{\mathcal{Z}}}(t_{i};\bar{\boldsymbol{Z}},t_{i+1}),t_{i}).
\end{align}
$\bar{\boldsymbol{\mathcal{Z}}}(t_{i};\bar{\boldsymbol{Z}},t_{i+1})$ denotes the phase space point at $t_{i}$, which passes through $\bar{\boldsymbol{Z}}$ at $t_{i+1}$ along the unperturbed orbit determined by Eq. \eqref{EqMotionofZbar}. $\delta \bar{f}(\bar{\boldsymbol{\mathcal{Z}}}(t_{i};\bar{\boldsymbol{Z}},t_{i+1}),t_{i})$ is the value of distribution function at the off-grid and is computed by the high-dimensional B-spline interpolation algorithm \cite{Xiao2017CCP}. 

After $\delta \bar{f}$ in the $\bar{\boldsymbol{Z}}$ coordinates is solved, $\delta f$ in the $\boldsymbol{Z}$ coordinates is solved by using the pull-back transform. According to Eq. \eqref{EqLieTrans} and the scalar invariance of the distribution function, $F(\boldsymbol{Z},t)=\bar{F}(\bar{\boldsymbol{Z}},t)$, the perturebed distribution function at $t_{i+1}$ is given by

\begin{align} \label{EqpullBack}
    \delta f = & \delta \bar{f} + \boldsymbol{G}_{1}\cdot \nabla(\bar{F}_{0}+\delta \bar{f}) 
    + \frac{1}{2} \boldsymbol{G}_{1}\cdot \nabla   \boldsymbol{G}_{1}\cdot \nabla(\bar{F}_{0}+\delta \bar{f})
\end{align}
The first-order generating vector $\boldsymbol{G}_{1}$ is computed by \eqref{EqG1}. In each given time interval $[t_{i},t_{i+1}]$, $S_{1}(\bar{\boldsymbol{Z}},t_{i})=0$ is used. $S_{1}(\bar{\boldsymbol{Z}},t_{i+1})$ is calculated by integrating Eq. \eqref{EqS1} along the unperturbed orbit from $t_{i}$ to $t_{i+1}$, which reads 

\begin{align} \label{EqComputS1}
    S_{1}(\bar{\boldsymbol{Z}},t_{i+1}) = \delta h(\bar{\boldsymbol{\mathcal{Z}}}(t_{i+1/2};\bar{\boldsymbol{Z}},t_{i+1}),t_{i+1/2})\Delta t.
\end{align}
Here, $\Delta t=t_{i+1}-t_{i}$ and $t_{i+1/2}=t_{i}+\frac{\Delta t}{2}$. In order to improve the computation accuracy, the midpoint prediction-correction algorithm has been applied in NLT \cite{Ye2016}.

Note that the effects of perturbed fields are taken into account by using the pull-back transform in the NLT code; the orbit within each $\Delta t$ is computed in the equilibrium fields. The equilibrium orbit does not change from time step to time step, and it can be computed in either the magnetic coordinates or the pseudo-Cartesian coordinates. Therefore, when including the magnetic axis, one computes the equilibrium orbit in the pseudo-Cartesian coordinates, and then map to the magnetic coordinates. 

The GK quasi-neutrality equation is solved by the usual finite-difference method \cite{DAI2019}, which will be introduced in Section \ref{SecAPPoisson}.

\subsection{Mean value theorem in field-alignd coordinates} \label{SecFAcoordinates} 
The field-aligned coordinates $\boldsymbol{X}=(\sqrt{\psi},\alpha,\theta)$ \cite{Hazeltine1988} are used in NLT \cite{XU2017}, with $\psi$ the poloidal magnetic flux, $\alpha = q(\psi)\theta-\zeta$ and $q$ the safety factor. The magnetic field in the field-aligned coordinates is generally written as $\boldsymbol{B}=\nabla \psi \times \nabla \alpha$ \cite{Hazeltine1988}. Therefore, $\nabla \alpha$ is perpendicular to the magnetic field line. The parallel gradient operator in the field-aligned coordinates is written as $\nabla_{\parallel}=\frac{\boldsymbol{B}\cdot\nabla}{B}=\frac{1}{J_{X}B}\partial_{\theta}$, with $J_{X}$ the space Jacobian. Drift waves vary slowly along the magnetic field. Therefore, the number of $\theta$ grid points is small in numerical computation, which greatly improves the computational efficiency.

When $r \to 0$, $\sqrt{\psi}=(\sqrt{\pi B_{0}/q_{0}})r$, with $q_{0}\equiv q(r=0)$. $\sqrt{\psi}$ can be understood as the usual minor radius. Therefore, $\sqrt{\psi}$ shall be understood as $r$. In the magnetic flux coordinates $(r,\theta,\zeta)$, a scalar function $\bar{h}(r,\theta,\zeta)$ is periodic in $\theta$ due to the single-value condition, with $r$ and $\zeta$ fixed. $\; \bar{\cdot} \;$ represents a scalar function in the magnetic flux coordinates. However, in the field-aligned coordinates $(r,\alpha,\theta)$, the scalar function $h(r,\theta,\zeta)$ is not periodic in $\theta$, with $r$ and $\alpha$ fixed. This can be seen from the transformation between $\bar{h}$ and $h$. For each toroidal Fourier component $h_n$, the transformation is given by

\begin{align} \label{EqAlnToFlux}
 \bar{h}_{n}(r,\theta)=  h_{n}(r,\theta) e^{inq(r)\theta},
\end{align}
where $ \bar{h}_{n}$ is periodic in  $\theta$. $h_{n}$ is not periodic in $\theta$, because $nq$ is not an integer. According to Eqs. \eqref{EqAxisValue} and \eqref{EqAlnToFlux}, the mean value theorem in the field-aligned coordinates is written as
\begin{align} \label{EqMeanValueFA}
      h_{n}(0,\theta_{k}) = 
      e^{-inq(0)\theta_{k}} \sum_{k=1}^{N_{\theta}} \bigg[\frac{4}{3N_{\theta}}h_{n}(\Delta r,\theta_k)e^{inq(\Delta r)\theta_k} - \notag \\
      \frac{1}{3N_{\theta}}h_{n}(2\Delta r,\theta_k)e^{inq(2\Delta r)\theta_k} \bigg],
\end{align}
Here, $h_n$ can be either the perturbed electrostatic potential $\delta \phi_{n}$ or the distribution function $\delta f_{n}$, since the dependence of the scalar function on the phase space coordinates will not affect our discussion.

\subsection{Application in solving the GK Vlasov equation in NLT} \label{SecAPVlasov}
In solving the GK Vlasov equation, the radial simulation domain is divided into two regions, $r>0$ and $r=0$. Firstly, $\delta f(r>0)$ is solved by using the characteristic line method and the numerical Lie transform \cite{Ye2016,XU2017}. Then $\delta f(r=0)$ is predicted by using $\delta f(r=\Delta r)$ and $\delta f(r=2\Delta r)$. Note that $\delta f$ is not periodic in $\theta$, therefore, $\delta f(r=0)$ is solved by predicting each toroidal mode $\delta f_{n}(r=0)$. Decomposing $\delta f$ into different toroidal modes gives

\begin{align} \label{EqdfN}
    \delta f(r,\alpha,\theta,v_{\parallel},\mu)=
    \sum_{n=-\frac{N_{\alpha}}{2}}^{\frac{N_{\alpha}}{2}} \delta f_{n} (r,\theta,v_{\parallel},\mu)e^{in\alpha}, 
\end{align}
with $N_{\alpha}$ the number of $\alpha$ grid points. Then, similar to Eq. \eqref{EqMeanValueFA}, the perturbed distribution function at the magnetic axis is calculated by 
\begin{align} \label{EqAxisDf}
      \delta f_{n}(0,\theta_{k},v_{\parallel},\mu) = 
      e^{-inq(0)\theta_{k}} \sum_{k=1}^{N_{\theta}} \bigg[\frac{4}{3N_{\theta}}\delta f_{n}(\Delta r,\theta_{k},v_{\parallel},\mu)e^{inq(\Delta r)\theta_k} - \notag \\
      \frac{1}{3N_{\theta}}\delta f_{n}(2\Delta r,\theta_{k},v_{\parallel},\mu)e^{inq(2\Delta r)\theta_k} \bigg],
\end{align}
Note that, in nonlinear simulation, the Fourier series expansions used in Eq. \eqref{EqdfN} has already been computed in the nonlinear filtering module. Therefore, to apply the mean value theorem, we just need to compute Eq. \eqref{EqAxisDf}. Previously, NLT \cite{DAI2019} solved $\delta f$ at the magnetic axis in the cylindrical coordinates to avoid Problem (i). One of the advantages of using the mean value theorem is that it provides the numerical solutions of the GK Vlasov equation and the GK quasi-neutrality equation,  with the numerical errors consistent with each other. The application of the generalized mean value theorem in solving GK Vlasov equation is shown in Appendix \ref{SecGeneralMeanGKV}.

\subsection{Application in solving the GK Poisson equation in NLT} \label{SecAPPoisson}
The perturbed electrostatic potential is decomposed into different toroidal modes in solving the GK quasi-neutrality equation, which reads

\begin{align} \label{EqdPhiN}
    \delta \phi (r,\alpha,\theta) = \sum_{n=-\frac{N_{\alpha}}{2}}^{\frac{N_{\alpha}}{2}} \delta \phi_{n} (r,\theta)e^{in\alpha},
\end{align}
In the ITG simulation, the maximum $m$ near the magnetic axis is dependent on the toroidal mode number $n$, since $m-nq$ is usually a small number. This indicates that high $n$ modes usually high $m$ components.

For $n \neq 0$ modes, the long-wavelength approximation is not applied to the GK quasi-neutrality equation, and $\delta \phi_{n\neq 0}$ is solved by Eq. \eqref{EqQn}, which is not a PDE. For $n\neq 0$, Eq. \eqref{EqQn} can be solved
iteratively, 

\begin{align} \label{EqQnNum}
    (c_{1}+c_{2})\delta \phi_{n}^{it+1}  = e_{i} \rho_{i,gy,n} + \int{\mathrm{d}^{3}v \left( \frac{c_{1}}{n_0}F \right) \left<\left< \delta \phi_{n}^{it} \right>\right>_{ga}},
\end{align}
where $it$ is the iteration number.

Although solving Eq. \eqref{EqQnNum} is untroubled by Problem (ii), the mean value method is used to achieve a numerical solution whose numerical error is consistent with $\delta f_{n}$. The application of the generalized mean value theorem in solving GK quasi-neutrality equation is shown in Appendix \ref{SecGeneralMeanGKP}. Similar to Eq. \eqref{EqMeanValueFA}, the electrostatic potential at the magnetic axis is predicted by
\begin{align} \label{EqAxisDphi}
      \delta\phi_{n}(0,\theta_{k}) = 
      e^{-inq(0)\theta_{k}} \sum_{k=1}^{N_{\theta}} \bigg[\frac{4}{3N_{\theta}}\delta\phi_{n}(\Delta r,\theta_k)e^{inq(\Delta r)\theta_k} - \notag \\
      \frac{1}{3N_{\theta}}\delta\phi_{n}(2\Delta r,\theta_k)e^{inq(2\Delta r)\theta_k} \bigg],
\end{align}

Therefore, in solving Eq. \eqref{EqQn} with $n\neq 0$, we compute $\delta \phi_{n}^{it+1}(r>0)$ by using Eq. \eqref{EqQnNum}, and predict $\delta \phi_{n}^{it+1}(r=0)$ by using Eq. \eqref{EqAxisDphi}.

For the $n=0$ mode, the long-wavelength approximation is usually satisfied. Therefore, $\delta \phi_{0}$ is solved by Eq. \eqref{EqPoisson}, the GK Poisson equation. The mean value theorem is used to provide the inner boundary condition at the magnetic axis. The outer boundary condition is given by 

\begin{align} \label{EqOutBoundN0}
    \delta \phi_{0}^{N_{r},k}=0, \quad k=1,2,\cdots,N_{\theta}, 
\end{align}
with $\phi_{0}^{j,k}=\phi_{0}(r_j,\theta_k)$. By integrating both sides of Eq. \eqref{EqPoisson} with $\int \mathrm{d}r \mathrm{d}\theta J_{X}$ and using the second-order central difference scheme, we obtain 

\begin{subequations} \label{EqPoissonDTs}
    \begin{align}
& \int_{V_{j,k}} \mathrm{d}r \mathrm{d}\theta J_{X} \nabla\cdot \left( c_{0} \nabla_{\perp}\delta\phi_{0} \right) = \sum_{j'=j-1}^{j+1} \sum_{k'=k-1}^{k+1} \alpha_{j,k}^{j',k'} \delta\phi_{0}^{j',k'}, \label{EqPoissonDTsA} \\
& \int_{V_{j,k}} \mathrm{d}r \mathrm{d}\theta J_{X} c_{1} \left( \delta\phi_{0} - \langle \delta \phi \rangle_{FA} \right) = \beta_{j,k}^{j,k} \delta\phi_{0}^{j,k} -  \sum_{k'=1}^{N_{\theta}} \nu_{j,k}^{j,k'} \delta\phi_{0}^{j,k'} ,\label{EqPoissonDTsB}  \\
& \int_{V_{j,k}} \mathrm{d}r \mathrm{d}\theta J_{X} e_{i}\rho_{i,gy,0} = \sigma_{j,k}^{j,k}\rho_{i,gy,0}^{j,k}, \label{EqPoissonDTsC}  
    \end{align}
\end{subequations}
where $V_{j,k}=[r_{j}- \frac{1}{2}\Delta r,r_{j}+\frac{1}{2}\Delta r] \times [\theta_{k}-\frac{1}{2}\Delta \theta, \theta_{k}+\frac{1}{2}\Delta \theta]$ is the integral domain. The coefficients $\alpha_{j,k}^{j',k'}$, $\beta_{j,k}^{j,k}$, $\nu_{j,k}^{j,k'}$, $\sigma_{j,k}^{j,k}$ are shown in Appendix \ref{SecCoef}. According to Eqs. \eqref{EqPoissonDTsA}-\eqref{EqPoissonDTsC}, the numerical equations at grid points $(r_j,\theta_{k})$ are given by 

\begin{align} \label{EqPoissonDEs}
  \sum_{j'=j-1}^{j+1} \sum_{k'=k-1}^{k+1} \alpha_{j,k}^{j',k'} \delta\phi_{0}^{j',k'}  - 
   \beta_{j,k}^{j,k} \delta\phi_{0}^{j,k} + \sum_{k'=1}^{N_{\theta}} \nu_{j,k}^{j,k'} \delta\phi_{0}^{j,k'} = 
   - \sigma_{j,k}^{j,k}\rho_{i,gy,0}^{j,k}, 
\end{align}
where the integer indices are given by 

\begin{align} 
& j = 2,3,\cdots,N_{r}-1  \nonumber\\
& k = 1,2,\cdots,N_{\theta}.
  \nonumber
\end{align}
Note that Eq. \eqref{EqPoissonDEs} is not defined at the magnetic axis. To obtain a unique solution (the number of equations is equal to that of variables), the numerical equations at the magnetic axis are provided by the mean value theorem, which reads 

\begin{align} \label{EqPoissonMA}
    \delta \phi_{0}^{1,k} = \frac{4}{3N_{\theta}}\sum_{k'=1}^{N_{\theta}} \delta \phi_{0}^{2,k'} - 
    \frac{1}{3N_{\theta}}\sum_{k'=1}^{N_{\theta}} \delta \phi_{0}^{3,k'}, \quad k=1,2,\cdots,N_{\theta}.
\end{align}
By combining Eqs. \eqref{EqOutBoundN0}, \eqref{EqPoissonDEs} and \eqref{EqPoissonMA}, we obtain full numerical equations to solve $\delta \phi_{0}$. Therefore, Eq. \eqref{EqPoissonMA} can be understood as an inner boundary condition.

\section{NLT simulation results including the magnetic axis} \label{sec4}
In this section, firstly, the validation of the proposed method is presented. Then the result of the GK simulation of the $n=0$ R-H test and the $n>0$ ITG mode is presented.

\subsection{Validation through the solution to the Poisson equation in the polar coordinates} \label{sec4A}
The GK Poisson equation is difficult to solve analytically in the magnetic flux coordinates. Therefore, we carry out the test in the polar coordinates $(r,\theta)$ to validate the Poisson equation solver using the mean value theorem. The Poisson equation in the polar coordinates is given by
\begin{align} \label{EqPoissonCyl}
    \frac{1}{r}\partial_r(r\partial_r\delta\phi)+\frac{1}{r^2}\partial^2_{\theta}\delta\phi = -e_{i}\rho_{i,gy} .
\end{align}
To test the Poisson equation solver, the perturbed electrostatic potential is given by $\delta\phi=\delta\phi_{0}(r)+\delta\phi_{1}(r)\cos{\theta}+\delta\phi_{2}(r)\cos{2\theta}$, with $\delta\phi_{0}=(1-r^2)e^{-25r^2}$, $\delta\phi_{1}=r(1-r^2)e^{-25r^2}$ and $\delta\phi_{2}=r^2(1-r^2)e^{-25r^2}$. Clearly, all the given poloidal Fourier components simultaneously satisfy: Eq. \eqref{EqSingleMSeries} near the magnetic axis, and the boundary condition $\delta\phi(r=1)=0$. The source term, $e_{i}\rho_{i,gy}$, is analytically calculated by using Eq. \eqref{EqPoissonCyl}. Then, the function $e_{i}\rho_{i,gy}$ is input into the Poisson equation solver to compute the numerical solution of $\delta\phi$.

\begin{figure}
\centering
\includegraphics[width=0.5\linewidth]{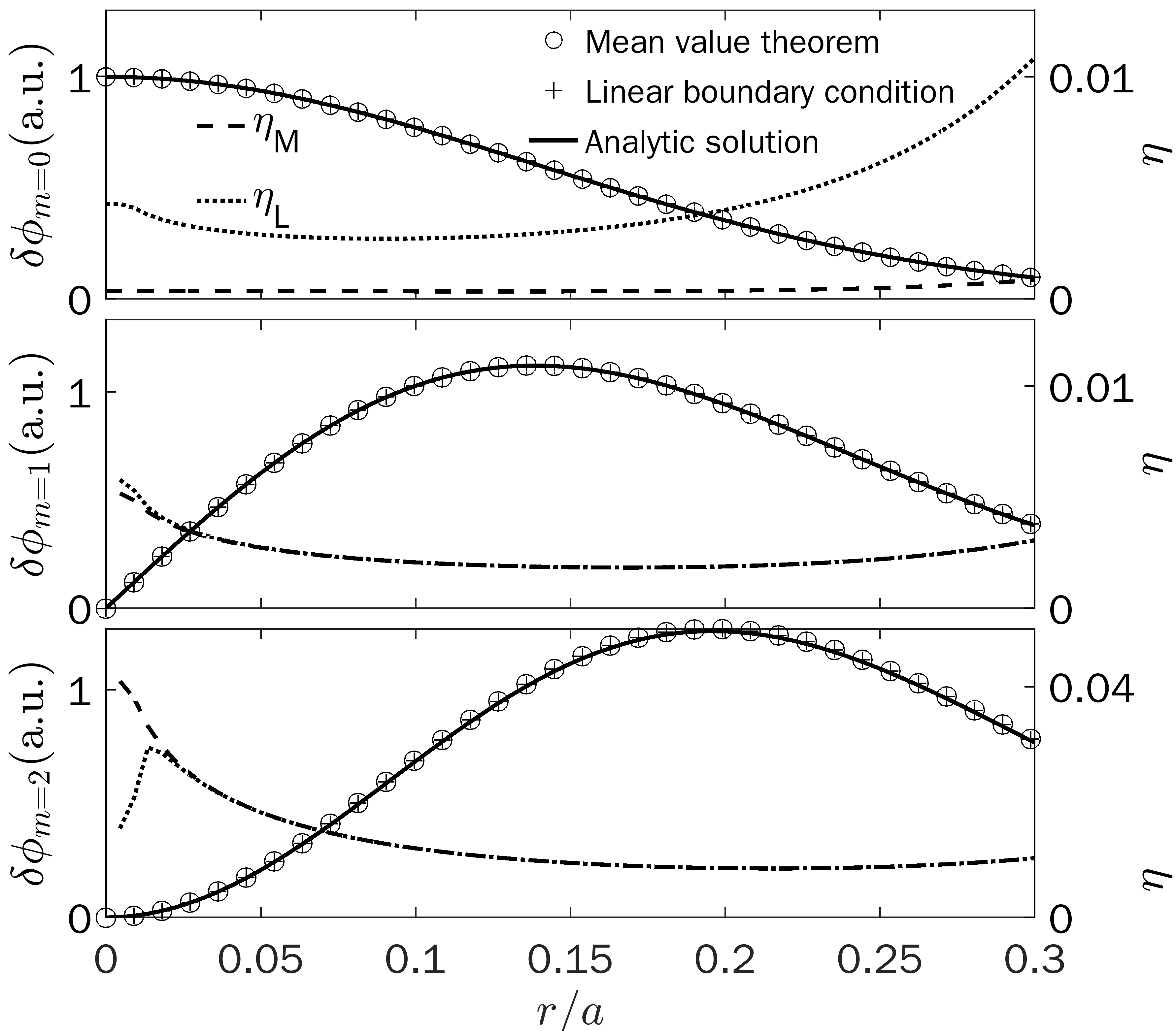}
\caption{\label{FigLaplacianTest} Numerical solutions to the Poisson equation constructed by respectively using the mean value theorem and the linear boundary conditions, compared with the analytic function for $\delta\phi$. $\eta_{M}$ and $\eta_{L}$ are the relative errors of the solutions obtained with the mean value theorem and the linear boundary condition, respectively.}
\end{figure}
The numerical solutions respectively using the mean value theorem and the linear boundary condition \cite{McClenaghan2014POP} are shown in Fig. \ref{FigLaplacianTest}. The linear boundary condition serves as an inner boundary condition for the finite-difference method, which reads $\delta\phi(r_{ib-1})=2\delta\phi(r_{ib})-\delta\phi(r_{ib+1})$. In Ref. \onlinecite{McClenaghan2014POP}, $ib=8$ was chosen for the Laplacian operator test. However, in our test, the larger $ib$ leads to more severe numerical error, due to the fact that $\partial^{2}_r\delta\phi(r_{ib})\neq 0$. Therefore, $ib=2$ is chosen in our test. Both numerical solutions exhibit similar relative errors for the $m=1$ and $m=2$ poloidal Fourier components when benchmarked against analytic solutions. The mean value theorem ensure the property of the solution that $\delta\phi_{m\neq 0}(r\to 0)\to0$. However, for the $m=0$ component, the relative error of the solution using the linear boundary condition is more severe than that of the solution using the mean value theorem; the more severe relative error arises because the given function of $\delta\phi_{0}$ does not satisfy the linear boundary condition. The linear boundary condition is first-order accurate in $r$, while the mean value theorem is second-order accurate in $r$, which is consistent with the second-order central difference scheme.

\subsection{R-H test}
The benchmark of R-H test away from the magnetic axis is shown in Appendix \ref{SecGammAwayAxis}. In this subsection, the R-H near the magnetic axis is carried out for the validation of the proposed method. According to Ref. \onlinecite{WANG2017}, an initial perturbed temperature will drive the electrostatic potential that balances it. This provides a convenient test for the radial force balance equation, which reads

\begin{align} \label{EqRadialforce}
  \delta E_{r} + \delta u_{\theta}B_{T}- \delta u_{\zeta}B_{P}-\frac{\delta p'_{i}}{n_{0}e_{i}}=0,
\end{align}
with $\delta E_{r}$ the perturbed radial electric field, $\delta u_{\theta}$ the perturbed poloidal flow, $\delta u_{\zeta}$ the perturbed toroidal flow, $B_{P}$ the poloidal magnetic field, $B_{T}$ the toroidal magnetic field, $\delta p_{i}=n_{0} \delta T_{i}$ and $\delta T_{i}$ the perturbed temperature. Here the prime represents the radial derivative. $\delta u_{\zeta}$, $\delta E_{r}$ and $\delta p_{i}$ are directly given by the simulation results, while the $\delta u_{\theta}$ is calculated from Eq. (\ref{EqRadialforce}). 

\begin{figure}
\centering
\includegraphics[width=0.5\linewidth]{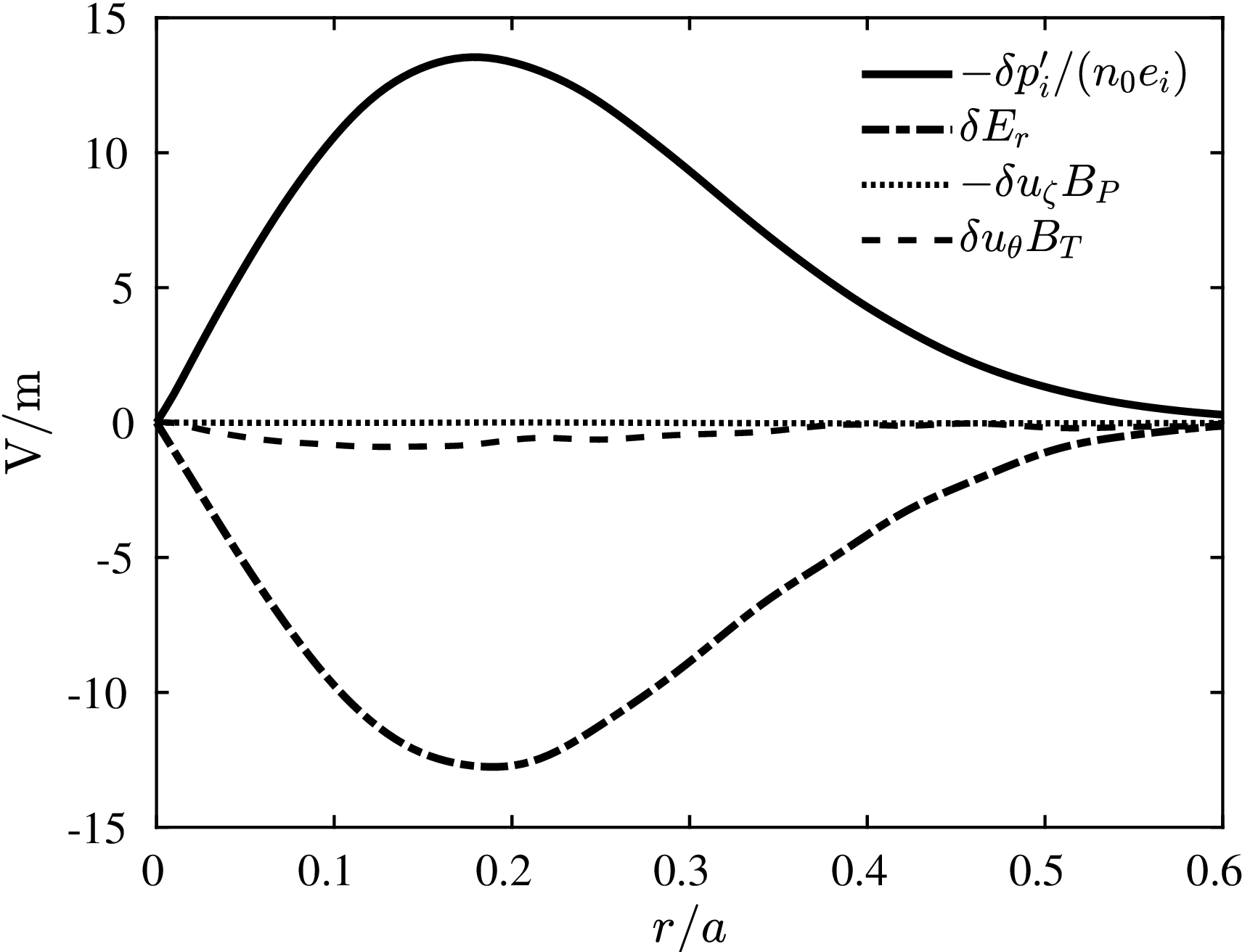}
\caption{\label{FigdTResi} Different terms in the radial force balance equation. }
\end{figure}

The simulation parameters are set as: $B_{0}=2.00\mathrm{T}$, the major radius $R_{0}=1.65\mathrm{m}$, the minor radius $a=0.40\mathrm{m}$. To avoid profile effects, the test is carried out in the radial homogeneous plasma, with equilibrium profiles $T_{0i}=1\mathrm{keV}$, $\tau_{e}\equiv T_{0e}/T_{0i}=1$, $n_{0}=10^{19}\mathrm{m}^{-3}$, $q=1.2$. The initial perturbed ion distribution function is given by

\begin{align} \label{EqRH2Pert}
  \delta f = \left( \frac{w}{T_{0i}} - \frac{3}{2} \right) \frac{\delta T_{i}}{T_{0i}}F_{0} ,
\end{align}
with $w$ the kinetic energy. Eq. \eqref{EqRH2Pert} gives the initial source as an ion heating impulse without density and  parallel momentum input. The initial perturbed temperature is set as

\begin{align} \label{EqRH2PertT}
  \frac{\delta T_{i}(r)}{T_{0i}} = 1.7 \times 10^{-3} \exp{ \left( - \frac{r^2}{\Delta_{\delta T}^2} \right) } ,
\end{align}
with $\Delta_{\delta T}=0.25a$. The $\Delta_{\delta T}$ is large enough to make the radial structure of $\delta T_{i}$ wider than the banana width of a trapped particle whose velocity approaching $v_{ti}$ near the magnetic axis. Although $T_{0i}$ is radial homogeneous, $v_{ti}$ is defined as $v_{ti}=\sqrt{2T_{0i}(r_0)/m_i}$, with $r_0=0$. The simulation results are shown in Fig. \ref{FigdTResi}. It can be seen that the pressure gradient is well balanced with the radial electric field.

\subsection{Linear ITG simulation}
\begin{figure}
\centering
\includegraphics[width=0.5\linewidth]{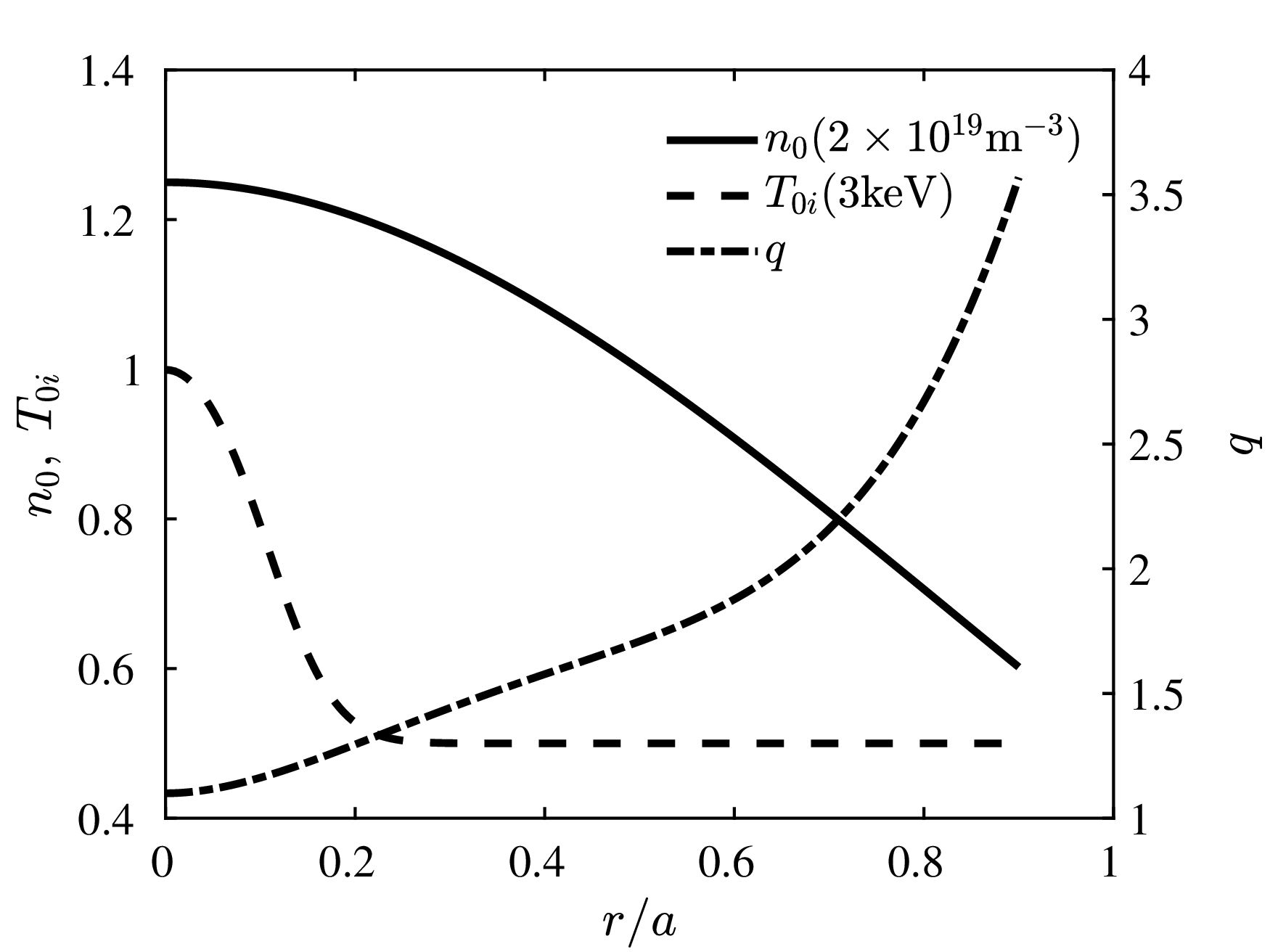}
\caption{\label{FigITG2Bal}  Equilibrium profiles used in the linear ITG simulation near the magnetic axis. }
\end{figure}

The benchmark of ITG simulation away from the magnetic axis is shown in Appendix \ref{SecITGAwayAixs}. In this subsection, the ITG simulation near the magnetic axis is carried out for the validation of the proposed method. The ITG mode tends to be more stable near the magnetic axis where the magnetic shear, $\hat{s}=\frac{q}{r}\frac{\mathrm{d}q}{\mathrm{d}r}$, is weaker. So, these tests are carried out with the internal transport barrier (ITB) like profiles for a relative high ITG growth rate near the magnetic axis. The main parameter are set as: $B_{0}=2.10\mathrm{T}$, $R_{0}=1.67\mathrm{m}$, $a=0.67\mathrm{m}$. Equilibrium profiles are based on the ITB data in DIII-D \cite{BURREL1998}. They are set as

\begin{subequations}
\begin{align}
    &q(r) = 1.10+7.79\left( \frac{r}{a} \right)^2 - 17.71\left( \frac{r}{a} \right)^3 + 13.46\left( \frac{r}{a} \right)^4, \\
    &n_{0i}(r) = 1.25 - 1.25\left( \frac{r}{a} \right)^2 + 0.50\left( \frac{r}{a} \right)^3, \\
    &T_{0i}(r) = -\frac{1}{2}\tanh{\left( \frac{r^{2}-r_{m}^{2}}{\Delta_{T}^2} \right)} +1,
\end{align}    
\end{subequations}
with $\Delta_{T}=0.15a$, $r_{m}=10^{-4}a$, $T_{0i}(r_0)=3\mathrm{keV}$ and $n_{0i}(r_0)=2\times10^{19}\mathrm{m^3}$. Details of equilibrium profiles are shown in Fig. \ref{FigITG2Bal}. Moreover, an equilibrium radial electric field $E_{0r}$ that balances the equilibrium pressure gradient is applied. The simulation domain are $r/a\in[0,0.9]$, $\theta\in[-\pi,\pi]$, $\alpha\in[0,2\pi]$, $v_{\parallel}/v_{ti}\in[3,3]$, $\mu B_{0}/T_{0i}(r_{0})=[0,9]$. 
Grid numbers are $\left(N_{r},N_{\alpha},N_{\theta},N_{v_\parallel},N_{\mu} \right)= (200,142,16,64,16)$. $\mu$ is discretized according to the Gauss-Legendre formula, while the other variables are discretized uniformly. \par

\begin{figure}
\centering
\includegraphics[width=0.5\linewidth]{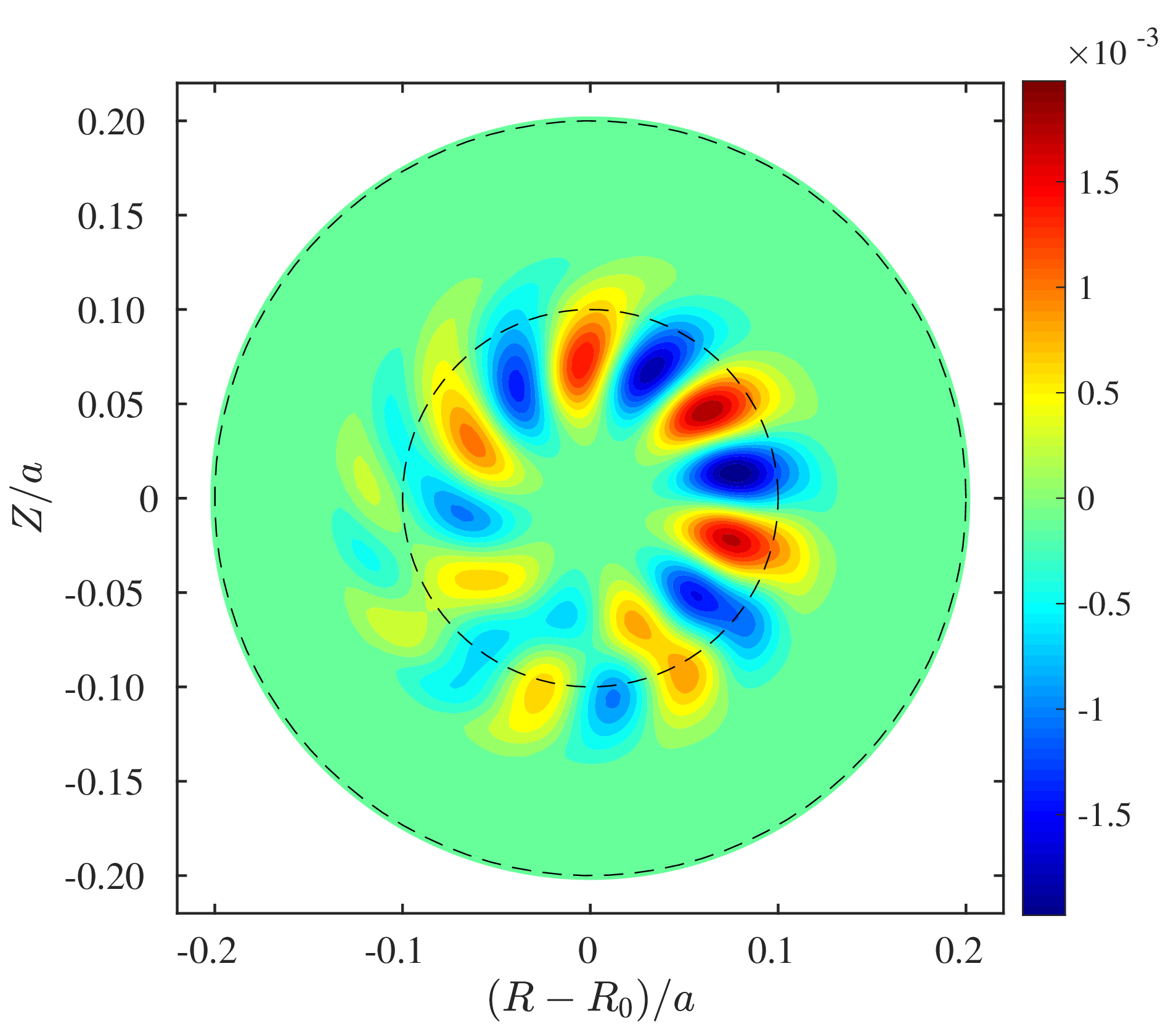}
\caption{\label{FigITG2Mode}  Mode structure of $\delta \phi$ with $n=6$. }
\end{figure}

The mode structure of the toroidal mode $n=6$, which is one of the most unstable modes, on a minor cross section is shown in Fig. \ref{FigITG2Mode}. It is a typical toroidal mode structure that balloons on the weak field side. On the strong field side, the balloon structure disappears and the amplitude of the mode is much weaker. 

As is discussed in Appendix \ref{SecITGwithGMeanVlaue}, near the magnetic axis, the numerical relative error of second-order central difference method becomes more severe (Problem (iii)) with the poloidal Fourier number, $m$, becomes larger. The generalized mean value theorem can be used to reduce the numerical error. However, the eigenvalues computed by using the mean value theorem and the generalized mean value theorem are almost the same, which is due to the fact that $\delta\phi_{n,m}\propto r^{|m|}$ goes more quickly to zero with larger $m$. This suggests that to solve Problem (iii) is a too strict requirement; the mean value theorem is good enough in practical applications.

\section{Summary and discussions} \label{sec5}
We have proposed a new computational method to solve the hyperbolic (such as the GK Vlasov) equation and the elliptic (such as the Poisson-like) equation based on the magnetic coordinates with the polar axis included. We have proved the mean value theorem, which indicates that the value of a scalar function at the polar axis can be predicted by the average of its neighbouring values, based on the continuity condition. This mean value theorem [Eq. \eqref{EqAxisValue}], which is understood as the discretization form of the continuity condition, systematically solves the pole problems including the problem of singular factor $1/r$ in the hyperbolic (GK Vlasov) equation and the problem of inner boundary condition in the elliptic (GK  Poisson) equation. The proposed method is used to update the NLT code to include the magnetic axis. 

The proposed method is validated by solving the GK Poisson equation (Section \ref{sec4A}). The NLT results of the $n=0$ R-H test near the magnetic axis agree well with the theoretical prediction. In the $n=6$ linear ITG simulation near the magnetic axis, a typical ballooning mode structure is found. 

It should be pointed out that the magnetic axis has been treated by using different computational methods. Problem (i) in the GK Poisson equation was usually solved in the cylindrical coordinates,
however Problem (ii) in the GK Poisson equation was solved in the magnetic (polar) coordinates; a coordinate transformation between cylindrical coordinates and magnetic surface coordinates is required to use these methods. The method proposed here solve both Problem (i) and Problem (ii) in the magnetic coordinates. For a global GK code based on the FDM, the proposed method can be easily used to update the code to include the magnetic axis.

\section*{Acknowledgement}
This work was supported by the National MCF Energy R\&D Program of China under Grant No. 2019YFE03060000, and the Strategic Priority Research Program of the Chinese Academy of Sciences under Grant Nos. XDB0790201 and XDB0500302.

\appendix

\section{Generalized mean value theorem} \label{SecGmeanValue}
The factor $r^{\lvert m \rvert}$ in Eq. \eqref{EqPolSeries} indicates that the numerical error of the low-order FDM near the pole may be serious; this is Problem (iii) in numerically solving PDEs at the pole, especially when there are high $m$ components in the system. 
To illustrate this problem, we evaluate the numerical error in using second-order central difference method to evaluate the $\partial_{r}\hat{g}_m$, with $\hat{g}_{m}(r)=r^{\lvert m \rvert}$.

\begin{align} \label{EqRadialDer}
 \frac{\partial \hat{g}_{m}}{\partial r} = \frac{(r_{j}+\Delta r)^{\lvert m\rvert}-(r_{j}-\Delta r)^{\lvert m\rvert}}{2\Delta r} \approx \lvert m \rvert r_{j}^{\lvert m \rvert - 1} + C_{\lvert m \rvert}^{3} \left( \frac{\Delta r}{r_{j}} \right)^{2}
 r_{j}^{\lvert m \rvert - 1}.
\end{align} 
The relative error is estimated to be 

\begin{align} \label{EqFDMError}
   \eta=\frac{(m-1)(m-2)}{3!}\left( \frac{\Delta r}{r_{j}} \right)^2,
\end{align}
which indicates that a larger $m$ will dramatically increase the $\eta$. Particularly, when $m>4$, the relative error at $r=r_2$ becomes $\eta>1$ and the numerical error is intolerable. To solve Problem (iii), one way is to use the Pade schemes \cite{LELE1992} to calculate the radial derivatives. However, the change from the explicit to the implicit scheme significantly reduces the computation efficiency by an order of $1/N_{r}$. 

\subsection{Numerical scheme}
Here we propose that the mean value theorem can be generalized to solve Problem (iii). 
Eq. \eqref{EqSingleMSeries} can be written as 

\begin{align} \label{EqMTrunc}
    \hat{g}_{m}(r) = r^{\lvert m \rvert } \left ( A_{m}^{(0)} + A_{m}^{(1)}r^{2} + \cdots \right ), 
\end{align}
which can be numerically evaluated at $r_{j_{-}}$ and $r_{j_{+}}$ ($j_{+}=j_{-}+1$) as

\begin{align}  \label{EqSolveCoeM}
\begin{cases}
    A_{m}^{(0)} r^{\lvert m \rvert}_{j_{-}} + A_{m}^{(1)} r^{\lvert m \rvert+2}_{j_{-}} = \hat{g}_{m}(r_{j_{-}}) , \\
    A_{m}^{(0)} r^{\lvert m \rvert}_{j_{+}} + A_{m}^{(1)} r^{\lvert m \rvert+2}_{j_{+}} = \hat{g}_{m}(r_{j_{+}}) .
\end{cases}
\end{align}
Using $\hat{g}(r_{j_{-}},\theta)$ and $\hat{g}(r_{j_{+}},\theta)$, one finds $\hat{g}_{m}(r_{-})$ and $\hat{g}_{m}(r_{+})$; using Eq. \eqref{EqSolveCoeM}, one finds the coefficients in Eq. \eqref{EqMTrunc}, which shall be used to predict $\hat{g}(r<r_{j_{-}},\theta)$. 
By writing $A_{m}^{(0)},A_{m}^{(1)}$ solved in Eq. \eqref{EqSolveCoeM} as $A_{m}^{(0)}(j_{-}),A_{m}^{(1)}(j_{-})$ respectively, one obtains

\begin{align} \label{EqGMVM}
    \hat{g}(r_j,\theta_{k}) = \sum_{m=-\frac{N_{\theta}}{2}}^{\frac{N_{\theta}}{2}} r_{j}^{\lvert m \rvert}\left[ A_{m}^{(0)}(j_{-}) +  A_{m}^{(1)}(j_{-}) \right]e^{im\theta_{k}}, \quad j<j_{-}.
\end{align}
This method can be understood as the "generalized mean value theorem", and the mean value theorem is the $j_{-}=2$ case.

To solve Problem (iii) by using the generalized mean value theorem, we solve the PDE by using the FDM when $r\geq r_{j_{-}}$, and the values of the function to be solved in the domain $r<r_{j_{-}}$ is predicted by Eq. \eqref{EqMTrunc}. According to Eq. \eqref{EqFDMError}, the numerical error of the low-order FDM decreases quickly with $r$ increasing. Note that the truncation error of Eq. \eqref{EqMTrunc} quickly decreases when approaching the pole.  Therefore, $j_{-}$ should be chosen to be large enough to keep a small FDM error and small enough to keep a small truncation error. In practice, for a system containing $m \leq 6$ components near the pole, $j_{-}=5$ can be chosen to find a good enough numerical solution, as will be discussed in Appendix \ref{SecITGwithGMeanVlaue}.

\subsection{Application in solving the GK Vlasov equation in NLT} \label{SecGeneralMeanGKV}
In solving the GK Vlasov equation, the radial simulation domain is divided into two regions, $r\geq r_{j_{}-}$ and $r< r_{j_{-}}$. The generalized mean value theorem is used in the region $r<r_{j_{-}}$.
Firstly, $\delta f(r\geq r_{j_{-}})$ is solved by using the characteristic line method and the numerical Lie transform \cite{Ye2016,XU2017}. Then $\delta f(r< r_{j_{-}})$ is predicted by using $\delta f(r=r_{j_{-}})$ and $\delta f(r=r_{j_{+}})$, with $j_{+}=j_{-}+1$. In the field-aligned coordinates, the range of $m$ is dependent on $n$, therefore, $\delta f(r<r_{j_{-}})$ is solved by predicting each toroidal mode $\delta f_{n}(r<r_{j_{-}})$. To calculate the coefficients for the prediction of $\delta f_{n}(r<r_{j_{-}})$. We need to know the values of all the poloidal Fourier components. By using the toroidal Fourier decomposition, Eq. \eqref{EqdfN}, and the coordinate transformation, Eq. \eqref{EqAlnToFlux}, the poloidal Fourier component of $\delta f_{n}$ is given by

\begin{align} \label{EqdfNM}
    \delta \bar{f}_{n,m}(r,v_{\parallel},\mu)= \frac{1}{2\pi}\int_{0}^{2\pi}\mathrm{d}\theta[\delta f_{n}(r,\theta,v_{\parallel},\mu)e^{inq\theta}]e^{-im\theta},
\end{align}
with $m_{0}-\frac{N_{\theta}}{2}\leq m\leq m_{0}+\frac{N_{\theta}}{2}$. Here, $m_{0} = [nq]$, with $[nq]$ the nearest integer around $nq$. According to Eq. \eqref{EqGMVM}, $\delta f_{n}(r<r_{j_{-}})$ is predicted by 

 \begin{align} \label{EqGMeanValueDf}
      \delta {f}_{n}(r_{j},\theta_{k},v_{\parallel},\mu) = e^{-inq(r_j)\theta_{k}}\sum_{m=m_{0}-\frac{N_{\theta}}{2}}^{m_{0}+\frac{N_{\theta}}{2}} r_{j}^{\lvert m \rvert} \left[ C_{n,m}^{(0)}(r_{j_{-}}) + C_{n,m}^{(1)}(r_{j_{-}})r_{j}^{2} \right]e^{im\theta_{k}}, & \notag \\ 
        j < j_{-}, & 
 \end{align}
where coefficients $C_{n,m}^{(0)}(r_{j_{-}}),C_{n,m}^{(1)}(r_{j_{-}})$ are calculated from Eq. \eqref{EqSolveCoeM} by replacing $\hat{g}_{m}(r_{j_{-}}), \hat{g}_{m}(r_{j_{+}})$  with $\delta\bar{f}_{n,m}(r_{j_{-}}), \delta\bar{f}_{n,m}(r_{j_{+}})$. After each $\delta f_n(r<r_{j_{-}})$ is solved through Eq. \eqref{EqGMeanValueDf}, $\delta f(r<r_{j_{-}})$ is calculated by using Eq. \eqref{EqdfN}. 

\subsection{Application in solving the GK quasi-neutrality equation in NLT} \label{SecGeneralMeanGKP}

The generalized mean value theorem can be used in solving $\delta\phi_{n}$, with $n\neq 0$, in order to achieve a numerical solution whose numerical error is consistent with $\delta f_{n}$. By using the toroidal Fourier decomposition, Eq. \eqref{EqdPhiN}, and the coordinate transformation, Eq. \eqref{EqAlnToFlux}, the poloidal poloidal Fourier component of $\delta \phi_{n}$ is given by

\begin{align} \label{EqdphiNM}
    \delta \bar{\phi}_{n,m}(r)=\frac{1}{2\pi}\int_{0}^{2\pi}\mathrm{d}\theta[\delta \phi_{n}(r,\theta)e^{inq\theta}]e^{-im\theta},
\end{align}
with $m_{0}-\frac{N_{\theta}}{2}\leq m\leq m_{0}+\frac{N_{\theta}}{2}$. Then $\delta \phi_{n}(r<r_{j_{-}})$ can be predicted by 

\begin{align} \label{EqGMTphi}
      \delta {\phi}_{n}(r_{j},\theta_{k}) = e^{-inq(r_j)\theta_{k}}\sum_{m=m_{0}-\frac{N_{\theta}}{2}}^{m_{0}+\frac{N_{\theta}}{2}} r_{j}^{\lvert m \rvert} \left[ D_{n,m}^{(0)}(r_{j_{-}}) + D_{n,m}^{(1)}(r_{j_{-}})r_{j}^{2} \right]e^{im\theta_{k}}, & \notag \\ 
        j < j_{-}, &     
\end{align}
where coefficients $D_{n,m}^{(0)}(r_{j_{-}}),D_{n,m}^{(1)}(r_{j_{-}})$ are calculated from Eq. \eqref{EqSolveCoeM} by replacing $\hat{g}_{m}(r_{j_{-}}), \hat{g}_{m}(r_{j_{+}})$  with $\delta\bar{\phi}_{n,m}(r_{j_{-}}), \delta\bar{\phi}_{n,m}(r_{j_{+}})$.

Therefore, in solving Eq. \eqref{EqQn} with the generalized mean value theorem, we compute $\delta \phi_{n}^{it+1}(r\geq r_{j_{-}})$ by using Eq. \eqref{EqQnNum}, and predict $\delta \phi_{n}^{it+1}(r < r_{j_{-}})$ by using Eq. \eqref{EqGMTphi}.

\subsection{Linear ITG simulation} \label{SecITGwithGMeanVlaue}
\begin{figure}
\centering
\includegraphics[width=0.5\linewidth]{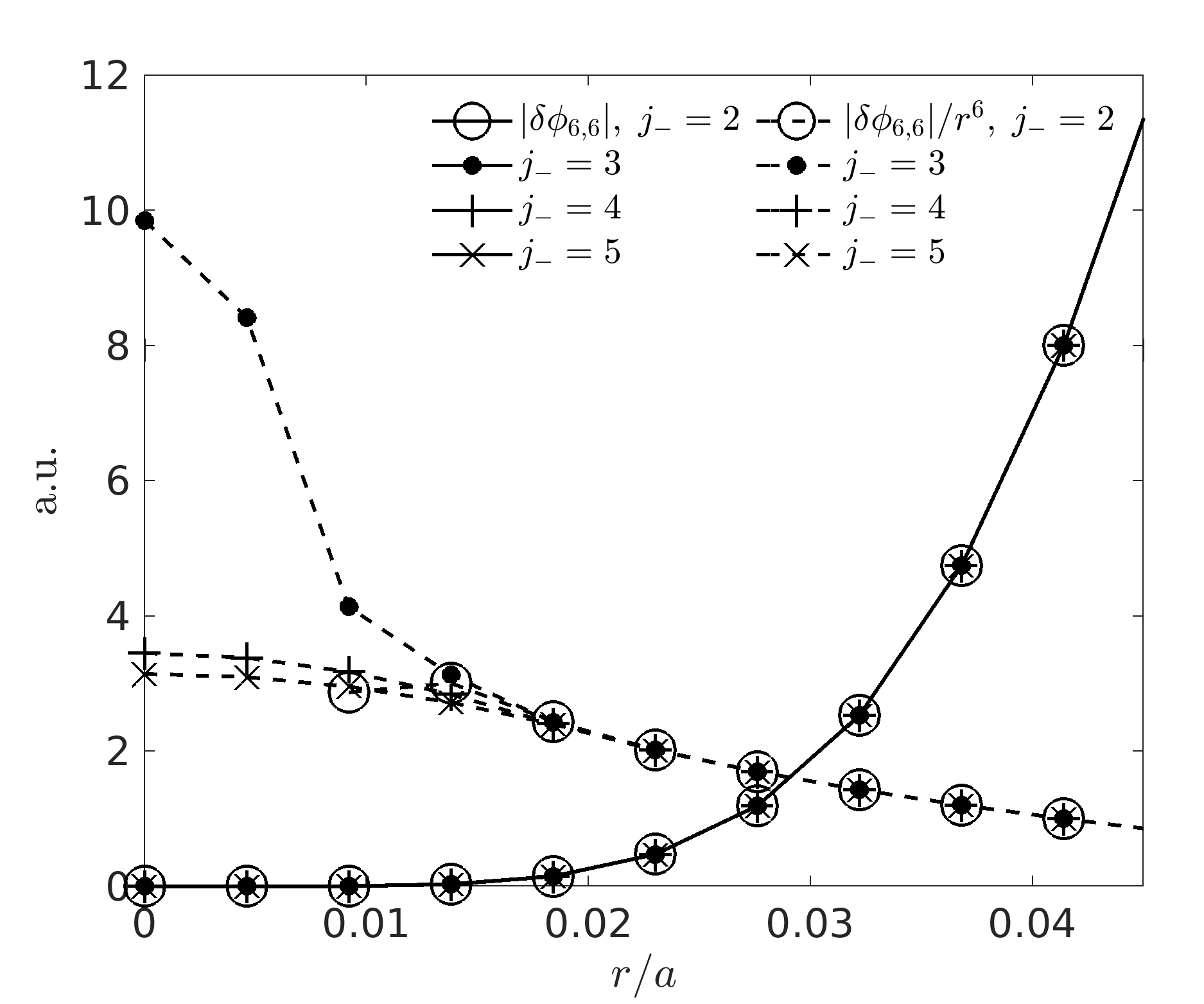}
\caption{\label{FigphimAxis} Eigenfunctions computed. Solid lines: $\lvert \delta \phi_{6,6} \rvert$; dashed lines: $\lvert \delta \phi_{6,6} \rvert/r^6$. $\lvert \delta \phi_{6,6} \rvert/r^6$ at $r_{1}$ and $r_{2}$ for $j_{-}=2$ are $466.6$ and $349.4$, which are not plotted here. }
\end{figure}

The equilibrium profiles used in this simulation are shown in Fig. \ref{FigITG2Bal}. The eigenfunctions of the $m/n=6/6$ harmonics $ \lvert \delta \phi_{6,6} \rvert$ for different $j_{-}$ near the magnetic axis are shown in Fig. \ref{FigphimAxis}. 
For $j_{-}=2,3,4,5$, $\lvert \delta \phi_{6,6} \rvert/r^{6}$ are convergent. The $\lvert \delta \phi_{6,6} \rvert/r^{6}$ for $j_{-}=2,3$ are not as good as those for $j_{-}=4,5$, which is consistent with the numerical relative error $\eta$ of the second-order central difference at $j_{-}$ shown in Table \ref{TabITG2Gamma}; $\eta$ for $j_{-}=2,3$ are much larger than those for $j_{-}=4,5$. However, the difference of $\lvert \delta \phi_{6,6} \rvert$ for different $j_{-}$ is not observable, since they are close to $0$ when approaching the magnetic axis; this can also be seen from $P_{j_{-}}$ shown in Table \ref{TabITG2Gamma}. For $j_{-}=2,3$, $\lvert \delta \phi_{6,6}(r_{j_{-}}) \rvert$ is smaller than $\lvert \delta \phi_{6,6}(r_{10}) \rvert$ by an order of $5\times 10^{-4}$. The linear growth rates and real frequencies for different $j_{-}$ shown in Table \ref{TabITG2Gamma} are almost the same, which is due to the reason that $\lvert \delta \phi_{6,6} \rvert$ are consistent with each other. This indicates that to have a small error of $\delta \phi_{n,m}/ r^{|m|}$ is a too strict requirement.

Clearly, to solve Problems (i) and (ii), one can use the mean value theorem [Eq. \eqref{EqAxisValue}]. The generalized mean value theorem [Eq. \eqref{EqGMVM}] can be used to solve Problem (iii); however, according to the above discussions, even with $j_{-}=2$, which corresponds to use the mean value theorem, the eigenvalue can be correctly computed, since the eigenfunction is nearly zero when approaching the magnetic axis. This suggests that to solve Problem (iii) is a too strict requirement. Therefore, one concludes that the mean value theorem is good enough in practical applications.

\begin{table}
\centering
\begin{tabular}{|c|c|c|c|c|}
\hline
$j_{-}$ & $\gamma\; (R_{0}/v_{ti})$ & $\omega\; (R_{0}/v_{ti})$ & $\eta$ at $r_{j_{-}}$ & $ P_{j_{-}}$ \\\hline
2 & 0.219 & $1.816$ & $333.3\%$ & $6.6 \times 10^{-4}$  \\
3 & 0.219 & $1.816$ & $83.3\%$  & $5.0 \times 10^{-4}$  \\
4 & 0.219 & $1.816$ & $37.0\%$  & $3.9 \times 10^{-3}$  \\
5 & 0.219 & $1.801$ & $20.8\%$  & $1.9 \times 10^{-2}$  \\\hline
\end{tabular}
\caption{\label{TabITG2Gamma} Linear growth rates,  real frequencies, the numerical relative error $\eta$ of the second-order central difference method at $r_{j_{-}}$, and $P_{j_{-}}= \delta \phi_{6,6}(r_{j_{-}})  /  \delta \phi_{6,6}(r_{10}) $, for different $j_{-}$.}
\end{table}

\section{R-H test away from the magnetic axis} \label{SecGammAwayAxis}

The R-H test of NLT has been performed in Ref. \onlinecite{Ye2016,DAI2019}. In this section, a R-H test away from the magnetic axis is performed as a benchmark for the NLT using the new computational method to treat the magnetic axis. Parameters are set as following: magnetic field at the axis $B_{0}=1.50\mathrm{T}$, major radius $R_{0}=1.25\mathrm{m}$, minor radius $a=0.45\mathrm{m}$. To avoid the phase mixing effect \cite{ZONCA2008}, the test is carried out in radial homogeneous plasma, with equilibrium profiles $q=1.2$, $T_{0i}=0.15\mathrm{keV}$, $\tau_{e}\equiv T_{0e}/T_{0i}=1$. The initial perturbation is given in the form of a radial perturbed density

\begin{align}  \label{EqGAM1perturb}
\frac{\delta n_{i}(r)}{n_{0}(r)} = 
\begin{cases}
    10^{-5} \sin{\left( \frac{r-r_a}{r_{b}-r_{a}} \right)}, \quad r\in\left[ r_a,r_b \right]  \\
    0, \qquad \qquad \qquad \quad \quad \mathrm{else} 
\end{cases}
\end{align}
with $\delta n_{i}$ the perturbed density, $r_a=0.4a$, $r_b=0.6a$. The radial simulation domain is $[0,0.85a]$, which particularly includes the magnetic axis. \par
\begin{figure}
\centering
\includegraphics[width=0.5\linewidth]{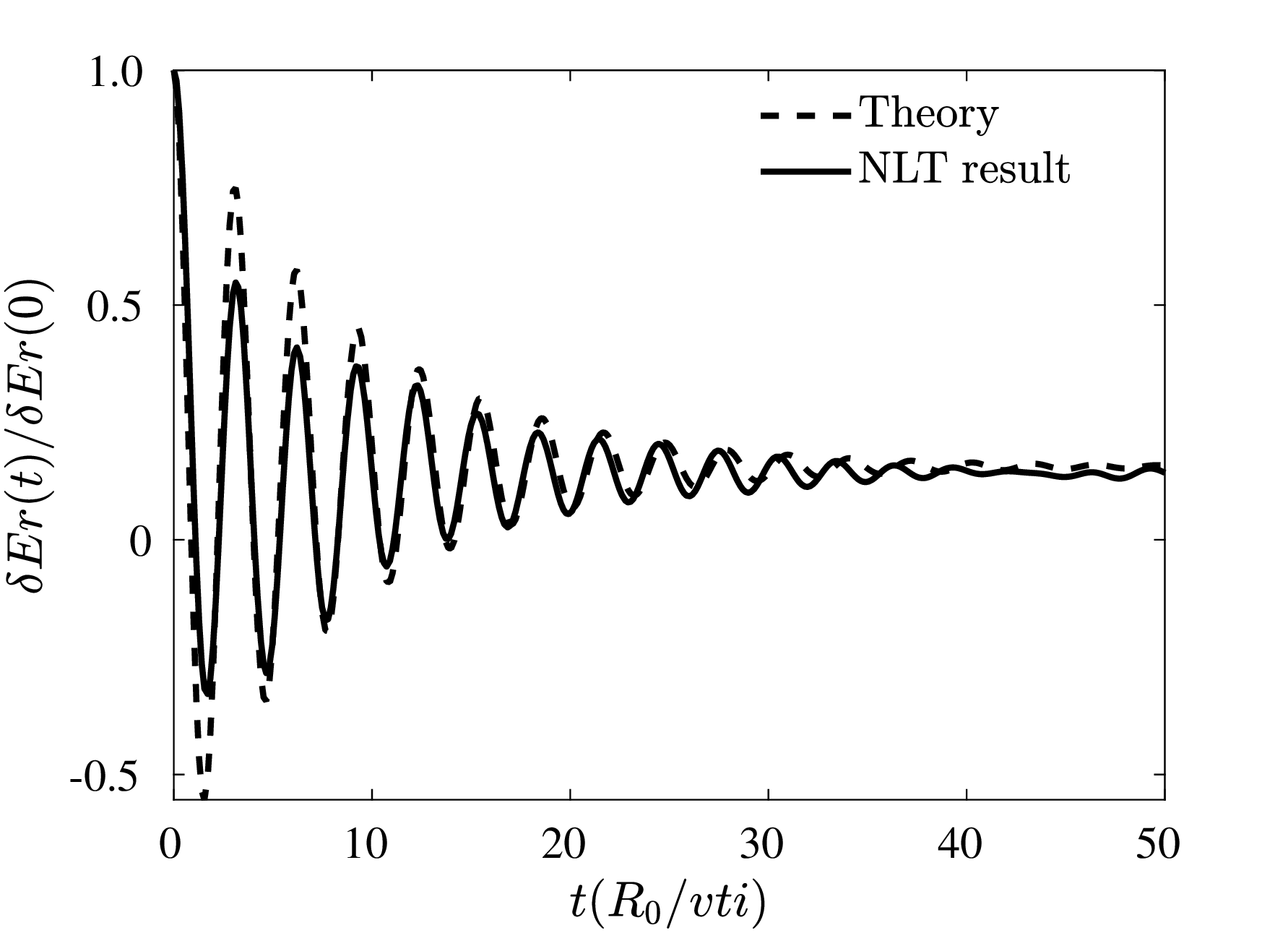}
\caption{\label{FigRHErt} Time evolution of perturbed radial electric field at $r_{0}$  }
\end{figure}

By taking account of the geodesic acoustic mode (GAM) oscillations, the collisionless damping and the residual flow \cite{ROUSENBLUTH1998}, the $m=n=0$ component of perturbed radial electric field is expected to behave as 

\begin{align} \label{EqGAM1Theory}
    \frac{\delta E_{r}(t)}{\delta E_{r}(0)} = R_{F} + (1-R_F)e^{-\gamma_{g} t}\cos{(\omega_{g}t)}
\end{align}
where $R_{F}=1/\left( 1+1.6q^{2}/\sqrt{\epsilon} \right)$ is the residual flow, $\epsilon=r/R_{0}$ is inverse aspect-ratio, $\omega_{g}$ and $\gamma_{g}$ are theoretical frequency and damping rate \cite{SUGAMA2006,Sugama2008}, respectively. Fig. \ref{FigRHErt} shows the time evolution of perturbed radial electric field at $r_{0}=0.5a$. The normalized unit of speed is defined as $v_{ti}=\sqrt{(2T_{0i}(r_0)/m_i)}$. The oscillation frequency, collisionless damping rate and residual flow all agree with the theoretical values.

\section{Linear ITG simulations away from the magnetic axis} \label{SecITGAwayAixs}

\begin{figure}
\centering
\includegraphics[width=0.8\linewidth]{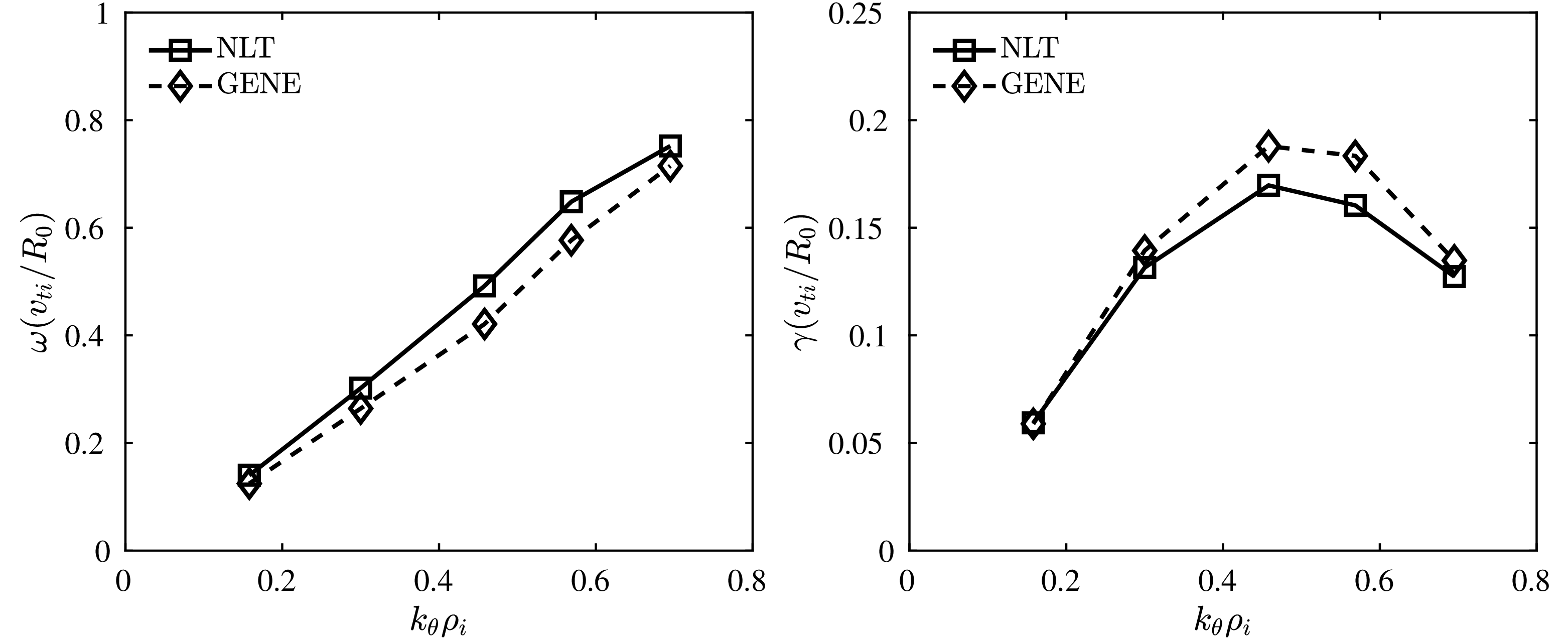}
\caption{\label{FigITG1Gamma} Comparison of linear ITG frequency (a) and growth rate (b) between GENE and NLT.  }
\end{figure}

The linear ITG tests of NLT have been performed in Ref. \onlinecite{Ye2016,DAI2019}. In this section, a set of linear ITG mode tests are performed to benchmark the NLT, which uses the new computational method to treat the magnetic axis, against another global GK code. These tests are carried out with the Cyclone Base Case (CBC) \cite{Dimits2000} parameters: $B_{0}=1.90\mathrm{T}$, $R_{0}=1.67\mathrm{m}$, $a=0.60\mathrm{m}$. The $q$ profile is set as

\begin{align} \label{EqITG1qPrf}
    q(r) = 0.854 + 2.4045 \left( \frac{r}{a} \right)^2
\end{align}
with $q_{0}\equiv q(r_0)=1.455$, $r_{0}=0.5$. The initial ion temperature and
density profile are set as 

\begin{align}  \label{EqITG1Profile}
  \hat{A}(r) = \frac{A(r)}{A(r_0)} =  \exp{ \left[ - \kappa_{A} \frac{a}{R_0} \Delta_{A} 
  \tanh{\left( \frac{r-r_{0}}{a} \right)}   \right] },
\end{align}
where $A$ can be chosen as either $T_{0i}$ or $n_{0i}$, and $T_{0i}(r_0) = 1.97\mathrm{keV}$, $n_{0i}=10^{19}\mathrm{m^3}$, $\Delta_{A}=0.30$, $\kappa_{n} \equiv R_{0}/L_{n}=2.23$, $\kappa_{T}=6.96$. $L_{n}$ and $L_{T}$ are the scale length of density and ion temperature, respectively. Here, $\tau_{e}=1$ is assumed. A comparison of linear ITG frequency and growth rate between different codes are shown in Fig. \ref{FigITG1Gamma}. The dimensionless number $k_{\theta}\rho_{i}$ is used to represent the toroidal mode number, where
$k_{\theta}$ is defined by $k_{\theta}=nq_{0}/r_{0}$. There are good agreements between simulation results of two codes.

\section{Coefficients in the discretized $n=0$ GK quasi-neutrality  equation} \label{SecCoef}
The coefficient $\alpha_{j,k}^{j',k'}$, with $j\geq 2$, is given by 

\begin{align} \label{EqCoefAlpha}
\alpha_{j,k}^{j',k'} = 
\begin{cases}
    \frac{1}{4}\Lambda_{r\theta}^{j+\frac{1}{2},k} + \frac{1}{4}\Lambda_{\theta r}^{j,k+\frac{1}{2}},  
    \quad  j'=j+1, k'=k+1, \\
    \quad \\
    \frac{\Delta \theta}{\Delta r} \Lambda_{rr}^{j+\frac{1}{2},k} + \frac{1}{4}\Lambda_{\theta r}^{j,k+\frac{1}{2}} - \frac{1}{4}\Lambda_{\theta r}^{j,k-\frac{1}{2}} 
    , \quad j'=j+1, k'=k, \\
    \quad \\
    -\frac{1}{4}\Lambda_{r\theta}^{j+\frac{1}{2},k} - \frac{1}{4}\Lambda_{\theta r}^{j,k-\frac{1}{2}}, 
    \quad j'=j+1, k'=k-1, \\    
    \quad \\
    \frac{1}{4}\Lambda_{r\theta}^{j+\frac{1}{2},k} - 
    \frac{1}{4}\Lambda_{r\theta}^{j-\frac{1}{2},k} +     \frac{\Delta r}{\Delta \theta}\Lambda_{\theta\theta}^{j,k+\frac{1}{2}},\quad j'=j,k'=k+1 \\
    \quad \\
    -\frac{\Delta\theta}{\Delta r}\Lambda_{rr}^{j+\frac{1}{2},k} - \frac{\Delta \theta}{\Delta r} \Lambda_{rr}^{j-\frac{1}{2},k}-\frac{\Delta r}{\Delta \theta}\Lambda_{\theta\theta}^{j,k+\frac{1}{2}}\\ -\frac{\Delta r}{\Delta \theta}\Lambda_{\theta\theta}^{j,k-\frac{1}{2}}, \quad j'=j,k'=k \\
    \quad \\
    -\frac{1}{4}\Lambda_{r\theta}^{j+\frac{1}{2},k} 
    + \frac{1}{4}\Lambda_{r\theta}^{j-\frac{1}{2},k} +\frac{\Delta r}{\Delta \theta}\Lambda_{\theta\theta}^{j,k-\frac{1}{2}}, \quad j'=j,k'=k-1 \\
    \quad \\
    -\frac{1}{4}\Lambda_{r\theta}^{j-\frac{1}{2},k} - \frac{1}{4}\Lambda_{r\theta}^{j,k+\frac{1}{2}} , \quad j'=j-1,k'=k+1\\
    \quad \\
    -\frac{\Delta \theta}{\Delta r} \Lambda_{rr}^{j-\frac{1}{2},k} - \frac{1}{4}\Lambda_{\theta r}^{j,k+\frac{1}{2}} + \frac{1}{4}\Lambda_{\theta r}^{j,k-\frac{1}{2}} 
    , \quad j'=j-1, k'=k, \\
    \quad \\
    \frac{1}{4}\Lambda_{r\theta}^{j-\frac{1}{2},k} + \frac{1}{4}\Lambda_{\theta r}^{j,k-\frac{1}{2}}, 
    \quad  j'=j-1, k'=k-1,   
\end{cases}
\end{align}
where $\Lambda_{ab}^{j,k}$ represents $\left[ c_{0}J_{X}g^{ab} \right]^{j,k}$ and $[\cdot]^{j,k}$ represents the value at the $(r_{j},\theta_{k})$. Here, the metric coefficient $g^{ab}$ is defined as 

\begin{align}
    g^{ab} = 
    \begin{cases}
        \nabla \theta \cdot \nabla \theta-\left( \frac{\partial_{r} \psi }{JB} \right)^2, \quad a=b=\theta \\
        \nabla a \cdot \nabla b, \quad \mathrm{else},
    \end{cases}
\end{align}
where $\psi$ is the poloidal magnetic flux. The coefficient $\beta_{j,k}^{j,k}$, with $j \geq 2$, is given by 

\begin{align} \label{EqCoefBeta}
    \beta_{j,k}^{j,k} = \Delta r\Delta \theta \left[ 
    c_{2}J_{X} \right]^{j,k}.
\end{align}
The coefficient $\nu_{j,k}^{j,k'}$, with $j \geq 2$, is given by 

\begin{align} \label{EqCoefNu}
    \nu_{j,k}^{j,k'} = \Delta r\Delta \theta \left[ 
    c_{2}J_{X} \right]^{j,k}\frac{\left[ J_{X} \right]^{j,k'}}{\sum_{k''=1}^{N_{\theta}}\left[ J_{X}\right]^{j,k''}}, \quad k'=1,2,\cdots,N_{\theta}.
\end{align}
The coefficient $\sigma_{j,k}^{j,k}$, with $j \geq 2$, is given by

\begin{align}
    \sigma_{j,k}^{j,k} = \Delta r\Delta \theta e_{i}\left[ J_{X} \right]^{j,k}.    
\end{align}

\nocite{*}
\bibliography{aipsamp}

\providecommand{\noopsort}[1]{}\providecommand{\singleletter}[1]{#1}%
\begin{thebibliography}{40}%
\makeatletter
\providecommand \@ifxundefined [1]{%
 \@ifx{#1\undefined}
}%
\providecommand \@ifnum [1]{%
 \ifnum #1\expandafter \@firstoftwo
 \else \expandafter \@secondoftwo
 \fi
}%
\providecommand \@ifx [1]{%
 \ifx #1\expandafter \@firstoftwo
 \else \expandafter \@secondoftwo
 \fi
}%
\providecommand \natexlab [1]{#1}%
\providecommand \enquote  [1]{``#1''}%
\providecommand \bibnamefont  [1]{#1}%
\providecommand \bibfnamefont [1]{#1}%
\providecommand \citenamefont [1]{#1}%
\providecommand \href@noop [0]{\@secondoftwo}%
\providecommand \href [0]{\begingroup \@sanitize@url \@href}%
\providecommand \@href[1]{\@@startlink{#1}\@@href}%
\providecommand \@@href[1]{\endgroup#1\@@endlink}%
\providecommand \@sanitize@url [0]{\catcode `\\12\catcode `\$12\catcode `\&12\catcode `\#12\catcode `\^12\catcode `\_12\catcode `\%12\relax}%
\providecommand \@@startlink[1]{}%
\providecommand \@@endlink[0]{}%
\providecommand \url  [0]{\begingroup\@sanitize@url \@url }%
\providecommand \@url [1]{\endgroup\@href {#1}{\urlprefix }}%
\providecommand \urlprefix  [0]{URL }%
\providecommand \Eprint [0]{\href }%
\providecommand \doibase [0]{http://dx.doi.org/}%
\providecommand \selectlanguage [0]{\@gobble}%
\providecommand \bibinfo  [0]{\@secondoftwo}%
\providecommand \bibfield  [0]{\@secondoftwo}%
\providecommand \translation [1]{[#1]}%
\providecommand \BibitemOpen [0]{}%
\providecommand \bibitemStop [0]{}%
\providecommand \bibitemNoStop [0]{.\EOS\space}%
\providecommand \EOS [0]{\spacefactor3000\relax}%
\providecommand \BibitemShut  [1]{\csname bibitem#1\endcsname}%
\let\auto@bib@innerbib\@empty
\bibitem [{\citenamefont {Constantinescu}\ and\ \citenamefont {Lele}(2002)}]{CONSTANTINESCU2002}%
  \BibitemOpen
  \bibfield  {author} {\bibinfo {author} {\bibfnamefont {G.~S.}\ \bibnamefont {Constantinescu}}\ and\ \bibinfo {author} {\bibfnamefont {S.~K.}\ \bibnamefont {Lele}},\ }\bibfield  {title} {\enquote {\bibinfo {title} {A highly accurate technique for the treatment of flow equations at the polar axis in cylindrical coordinates using series expansions},}\ }\href@noop {} {\bibfield  {journal} {\bibinfo  {journal} {J. Comput. Phys.}\ }\textbf {\bibinfo {volume} {183}},\ \bibinfo {pages} {165} (\bibinfo {year} {2002})}\BibitemShut {NoStop}%
\bibitem [{\citenamefont {Griffin}, \citenamefont {Jones},\ and\ \citenamefont {Anderson}(1979)}]{GRIFFIN1979}%
  \BibitemOpen
  \bibfield  {author} {\bibinfo {author} {\bibfnamefont {M.~D.}\ \bibnamefont {Griffin}}, \bibinfo {author} {\bibfnamefont {E.}~\bibnamefont {Jones}}, \ and\ \bibinfo {author} {\bibfnamefont {J.~D.}\ \bibnamefont {Anderson}},\ }\bibfield  {title} {\enquote {\bibinfo {title} {A computational fluid dynamic technique valid at the centerline for non-axisymmetric problems in cylindrical coordinates},}\ }\href@noop {} {\bibfield  {journal} {\bibinfo  {journal} {J. Comput. Phys.}\ }\textbf {\bibinfo {volume} {30}},\ \bibinfo {pages} {352} (\bibinfo {year} {1979})}\BibitemShut {NoStop}%
\bibitem [{\citenamefont {Huang}\ and\ \citenamefont {Sloan}(1993)}]{HUANG1993}%
  \BibitemOpen
  \bibfield  {author} {\bibinfo {author} {\bibfnamefont {W.}~\bibnamefont {Huang}}\ and\ \bibinfo {author} {\bibfnamefont {D.~M.}\ \bibnamefont {Sloan}},\ }\bibfield  {title} {\enquote {\bibinfo {title} {Pole condition for singular problems: The pseudospectral approximation},}\ }\href@noop {} {\bibfield  {journal} {\bibinfo  {journal} {J. Comput. Phys.}\ }\textbf {\bibinfo {volume} {107}},\ \bibinfo {pages} {254} (\bibinfo {year} {1993})}\BibitemShut {NoStop}%
\bibitem [{\citenamefont {Matsushima}\ and\ \citenamefont {Marcus}(1995)}]{MATSUSHIMA1995}%
  \BibitemOpen
  \bibfield  {author} {\bibinfo {author} {\bibfnamefont {T.}~\bibnamefont {Matsushima}}\ and\ \bibinfo {author} {\bibfnamefont {P.~S.}\ \bibnamefont {Marcus}},\ }\bibfield  {title} {\enquote {\bibinfo {title} {A spectral method for polar coordinates},}\ }\href@noop {} {\bibfield  {journal} {\bibinfo  {journal} {J. Comput. Phys.}\ }\textbf {\bibinfo {volume} {120}},\ \bibinfo {pages} {365} (\bibinfo {year} {1995})}\BibitemShut {NoStop}%
\bibitem [{\citenamefont {Mohseni}\ and\ \citenamefont {Colonius}(2000)}]{MOHSENI2000787}%
  \BibitemOpen
  \bibfield  {author} {\bibinfo {author} {\bibfnamefont {K.}~\bibnamefont {Mohseni}}\ and\ \bibinfo {author} {\bibfnamefont {T.}~\bibnamefont {Colonius}},\ }\bibfield  {title} {\enquote {\bibinfo {title} {Numerical treatment of polar coordinate singularities},}\ }\href@noop {} {\bibfield  {journal} {\bibinfo  {journal} {J. Comput. Phys.}\ }\textbf {\bibinfo {volume} {157}},\ \bibinfo {pages} {787} (\bibinfo {year} {2000})}\BibitemShut {NoStop}%
\bibitem [{\citenamefont {McClenaghan}\ \emph {et~al.}(2014)\citenamefont {McClenaghan}, \citenamefont {Lin}, \citenamefont {Holod}, \citenamefont {Deng},\ and\ \citenamefont {Wang}}]{McClenaghan2014POP}%
  \BibitemOpen
  \bibfield  {author} {\bibinfo {author} {\bibfnamefont {J.}~\bibnamefont {McClenaghan}}, \bibinfo {author} {\bibfnamefont {Z.}~\bibnamefont {Lin}}, \bibinfo {author} {\bibfnamefont {I.}~\bibnamefont {Holod}}, \bibinfo {author} {\bibfnamefont {W.}~\bibnamefont {Deng}}, \ and\ \bibinfo {author} {\bibfnamefont {Z.}~\bibnamefont {Wang}},\ }\bibfield  {title} {\enquote {\bibinfo {title} {{Verification of gyrokinetic particle simulation of currentdriven instability in fusion plasmas. I. Internal kink mode}},}\ }\href@noop {} {\bibfield  {journal} {\bibinfo  {journal} {Phys. Plasmas}\ }\textbf {\bibinfo {volume} {21}},\ \bibinfo {pages} {122519} (\bibinfo {year} {2014})}\BibitemShut {NoStop}%
\bibitem [{\citenamefont {Bouzat}\ \emph {et~al.}(2018)\citenamefont {Bouzat}, \citenamefont {Bressan}, \citenamefont {Grandgirard}, \citenamefont {Latu},\ and\ \citenamefont {Mehrenberger}}]{Bouzat2018ESAIM}%
  \BibitemOpen
  \bibfield  {author} {\bibinfo {author} {\bibfnamefont {N.}~\bibnamefont {Bouzat}}, \bibinfo {author} {\bibfnamefont {C.}~\bibnamefont {Bressan}}, \bibinfo {author} {\bibfnamefont {V.}~\bibnamefont {Grandgirard}}, \bibinfo {author} {\bibfnamefont {G.}~\bibnamefont {Latu}}, \ and\ \bibinfo {author} {\bibfnamefont {M.}~\bibnamefont {Mehrenberger}},\ }\bibfield  {title} {\enquote {\bibinfo {title} {{TARGETING REALISTIC GEOMETRY IN TOKAMAK CODE GYSELA}},}\ }\href@noop {} {\bibfield  {journal} {\bibinfo  {journal} {ESAIM: Proceedings}\ }\textbf {\bibinfo {volume} {63}},\ \bibinfo {pages} {179} (\bibinfo {year} {2018})}\BibitemShut {NoStop}%
\bibitem [{\citenamefont {Wang}, \citenamefont {Wang},\ and\ \citenamefont {Wu}(2024)}]{WANG2023}%
  \BibitemOpen
  \bibfield  {author} {\bibinfo {author} {\bibfnamefont {S.}~\bibnamefont {Wang}}, \bibinfo {author} {\bibfnamefont {Z.}~\bibnamefont {Wang}}, \ and\ \bibinfo {author} {\bibfnamefont {T.}~\bibnamefont {Wu}},\ }\bibfield  {title} {\enquote {\bibinfo {title} {Self-organized evolution of the internal transport barrier in ion-temperature-gradient driven gyrokinetic turbulence},}\ }\href@noop {} {\bibfield  {journal} {\bibinfo  {journal} {Phys. Rev. Lett.}\ }\textbf {\bibinfo {volume} {132}},\ \bibinfo {pages} {065106} (\bibinfo {year} {2024})}\BibitemShut {NoStop}%
\bibitem [{\citenamefont {Brizard}\ and\ \citenamefont {Hahm}(2007)}]{BRIZARD2007}%
  \BibitemOpen
  \bibfield  {author} {\bibinfo {author} {\bibfnamefont {A.~J.}\ \bibnamefont {Brizard}}\ and\ \bibinfo {author} {\bibfnamefont {T.~S.}\ \bibnamefont {Hahm}},\ }\bibfield  {title} {\enquote {\bibinfo {title} {Foundations of nonlinear gyrokinetic theory},}\ }\href@noop {} {\bibfield  {journal} {\bibinfo  {journal} {Rev. Mod. Phys.}\ }\textbf {\bibinfo {volume} {79}},\ \bibinfo {pages} {421} (\bibinfo {year} {2007})}\BibitemShut {NoStop}%
\bibitem [{\citenamefont {Lee}(1983)}]{LEE1983}%
  \BibitemOpen
  \bibfield  {author} {\bibinfo {author} {\bibfnamefont {W.~W.}\ \bibnamefont {Lee}},\ }\bibfield  {title} {\enquote {\bibinfo {title} {{Gyrokinetic approach in particle simulation}},}\ }\href@noop {} {\bibfield  {journal} {\bibinfo  {journal} {Phys. Fluids}\ }\textbf {\bibinfo {volume} {26}},\ \bibinfo {pages} {556} (\bibinfo {year} {1983})}\BibitemShut {NoStop}%
\bibitem [{\citenamefont {Jolliet}(2009)}]{Jolliet2009}%
  \BibitemOpen
  \bibfield  {author} {\bibinfo {author} {\bibfnamefont {S.}~\bibnamefont {Jolliet}},\ }\bibfield  {title} {\enquote {\bibinfo {title} {{Gyrokinetic Particle-in-Cell Global Simulations of IonTemperature-Gradient and Collisionless-Trapped-Electron-Mode Turbulence in Tokamaks}},}\ }\href@noop {} {\bibfield  {journal} {\bibinfo  {journal} {\'Ecole Polytechnique F\'ed\'erale de Lausanne}\ } (\bibinfo {year} {2009})}\BibitemShut {NoStop}%
\bibitem [{\citenamefont {Lapillon}(2010)}]{Lapillon2010}%
  \BibitemOpen
  \bibfield  {author} {\bibinfo {author} {\bibfnamefont {X.}~\bibnamefont {Lapillon}},\ }\bibfield  {title} {\enquote {\bibinfo {title} {{Local and Global Eulerian Gyrokinetic Simulations of Microturbulence in Realistic Geometry with Applications to the TCV Tokamak}},}\ }\href@noop {} {\bibfield  {journal} {\bibinfo  {journal} {\'Ecole Polytechnique F\'ed\'erale de Lausanne}\ } (\bibinfo {year} {2010})}\BibitemShut {NoStop}%
\bibitem [{\citenamefont {Lanti}\ \emph {et~al.}(2020)\citenamefont {Lanti}, \citenamefont {Ohana}, \citenamefont {Tronko}, \citenamefont {Hayward-Schneider}, \citenamefont {Bottino}, \citenamefont {McMillan}, \citenamefont {Mishchenko}, \citenamefont {Scheinberg}, \citenamefont {Biancalani}, \citenamefont {Angelino}, \citenamefont {Brunner}, \citenamefont {Dominski}, \citenamefont {Donnel}, \citenamefont {Gheller}, \citenamefont {Hatzky}, \citenamefont {Jocksch}, \citenamefont {Jolliet}, \citenamefont {Lu}, \citenamefont {Martin~Collar}, \citenamefont {Novikau}, \citenamefont {Sonnendr\"ucker}, \citenamefont {Vernay},\ and\ \citenamefont {Villard}}]{Lanti2019}%
  \BibitemOpen
  \bibfield  {author} {\bibinfo {author} {\bibfnamefont {E.}~\bibnamefont {Lanti}}, \bibinfo {author} {\bibfnamefont {N.}~\bibnamefont {Ohana}}, \bibinfo {author} {\bibfnamefont {N.}~\bibnamefont {Tronko}}, \bibinfo {author} {\bibfnamefont {T.}~\bibnamefont {Hayward-Schneider}}, \bibinfo {author} {\bibfnamefont {A.}~\bibnamefont {Bottino}}, \bibinfo {author} {\bibfnamefont {B.~F.}\ \bibnamefont {McMillan}}, \bibinfo {author} {\bibfnamefont {A.}~\bibnamefont {Mishchenko}}, \bibinfo {author} {\bibfnamefont {A.}~\bibnamefont {Scheinberg}}, \bibinfo {author} {\bibfnamefont {A.}~\bibnamefont {Biancalani}}, \bibinfo {author} {\bibfnamefont {P.}~\bibnamefont {Angelino}}, \bibinfo {author} {\bibfnamefont {S.}~\bibnamefont {Brunner}}, \bibinfo {author} {\bibfnamefont {J.}~\bibnamefont {Dominski}}, \bibinfo {author} {\bibfnamefont {P.}~\bibnamefont {Donnel}}, \bibinfo {author} {\bibfnamefont {C.}~\bibnamefont {Gheller}}, \bibinfo {author} {\bibfnamefont {R.}~\bibnamefont {Hatzky}}, \bibinfo {author} {\bibfnamefont
  {A.}~\bibnamefont {Jocksch}}, \bibinfo {author} {\bibfnamefont {S.}~\bibnamefont {Jolliet}}, \bibinfo {author} {\bibfnamefont {Z.~X.}\ \bibnamefont {Lu}}, \bibinfo {author} {\bibfnamefont {J.~P.}\ \bibnamefont {Martin~Collar}}, \bibinfo {author} {\bibfnamefont {I.}~\bibnamefont {Novikau}}, \bibinfo {author} {\bibfnamefont {E.}~\bibnamefont {Sonnendr\"ucker}}, \bibinfo {author} {\bibfnamefont {T.}~\bibnamefont {Vernay}}, \ and\ \bibinfo {author} {\bibfnamefont {L.}~\bibnamefont {Villard}},\ }\bibfield  {title} {\enquote {\bibinfo {title} {{ORB5: A global electromagnetic gyrokinetic code using the PIC approach in toroidal geometry}},}\ }\href@noop {} {\bibfield  {journal} {\bibinfo  {journal} {Comput. Phys. Comm.}\ }\textbf {\bibinfo {volume} {251}},\ \bibinfo {pages} {107072} (\bibinfo {year} {2020})}\BibitemShut {NoStop}%
\bibitem [{\citenamefont {Lin}\ \emph {et~al.}(1998)\citenamefont {Lin}, \citenamefont {Hahm}, \citenamefont {Lee}, \citenamefont {Tang},\ and\ \citenamefont {White}}]{LIN1998}%
  \BibitemOpen
  \bibfield  {author} {\bibinfo {author} {\bibfnamefont {Z.}~\bibnamefont {Lin}}, \bibinfo {author} {\bibfnamefont {T.~S.}\ \bibnamefont {Hahm}}, \bibinfo {author} {\bibfnamefont {W.~W.}\ \bibnamefont {Lee}}, \bibinfo {author} {\bibfnamefont {W.~M.}\ \bibnamefont {Tang}}, \ and\ \bibinfo {author} {\bibfnamefont {R.~B.}\ \bibnamefont {White}},\ }\bibfield  {title} {\enquote {\bibinfo {title} {Turbulent transport reduction by zonal flows: Massively parallel simulations},}\ }\href@noop {} {\bibfield  {journal} {\bibinfo  {journal} {Science}\ }\textbf {\bibinfo {volume} {281}},\ \bibinfo {pages} {1835} (\bibinfo {year} {1998})}\BibitemShut {NoStop}%
\bibitem [{\citenamefont {Candy}\ and\ \citenamefont {Waltz}(2003)}]{CANDY2003}%
  \BibitemOpen
  \bibfield  {author} {\bibinfo {author} {\bibfnamefont {J.}~\bibnamefont {Candy}}\ and\ \bibinfo {author} {\bibfnamefont {R.~E.}\ \bibnamefont {Waltz}},\ }\bibfield  {title} {\enquote {\bibinfo {title} {An eulerian gyrokinetic-maxwell solver},}\ }\href@noop {} {\bibfield  {journal} {\bibinfo  {journal} {J. Comput. Phys.}\ }\textbf {\bibinfo {volume} {186}},\ \bibinfo {pages} {545} (\bibinfo {year} {2003})}\BibitemShut {NoStop}%
\bibitem [{\citenamefont {Grandgirard}\ \emph {et~al.}(2006)\citenamefont {Grandgirard}, \citenamefont {Sarazin}, \citenamefont {Garbet}, \citenamefont {Dif‐Pradalier}, \citenamefont {Ghendrih}, \citenamefont {Crouseilles}, \citenamefont {Latu}, \citenamefont {Sonnendrücker}, \citenamefont {Besse},\ and\ \citenamefont {Bertrand}}]{Grandgirard2006}%
  \BibitemOpen
  \bibfield  {author} {\bibinfo {author} {\bibfnamefont {V.}~\bibnamefont {Grandgirard}}, \bibinfo {author} {\bibfnamefont {Y.}~\bibnamefont {Sarazin}}, \bibinfo {author} {\bibfnamefont {X.}~\bibnamefont {Garbet}}, \bibinfo {author} {\bibfnamefont {G.}~\bibnamefont {Dif‐Pradalier}}, \bibinfo {author} {\bibfnamefont {P.}~\bibnamefont {Ghendrih}}, \bibinfo {author} {\bibfnamefont {N.}~\bibnamefont {Crouseilles}}, \bibinfo {author} {\bibfnamefont {G.}~\bibnamefont {Latu}}, \bibinfo {author} {\bibfnamefont {E.}~\bibnamefont {Sonnendrücker}}, \bibinfo {author} {\bibfnamefont {N.}~\bibnamefont {Besse}}, \ and\ \bibinfo {author} {\bibfnamefont {P.}~\bibnamefont {Bertrand}},\ }\bibfield  {title} {\enquote {\bibinfo {title} {{GYSELA, a full‐f global gyrokinetic Semi‐Lagrangian code for ITG turbulence simulations}},}\ }\href@noop {} {\bibfield  {journal} {\bibinfo  {journal} {AIP Conf. Proc.}\ }\textbf {\bibinfo {volume} {871}},\ \bibinfo {pages} {100} (\bibinfo {year} {2006})}\BibitemShut {NoStop}%
\bibitem [{\citenamefont {Jolliet}\ \emph {et~al.}(2007)\citenamefont {Jolliet}, \citenamefont {Bottino}, \citenamefont {Angelino}, \citenamefont {Hatzky}, \citenamefont {Tran}, \citenamefont {Mcmillan}, \citenamefont {Sauter}, \citenamefont {Appert}, \citenamefont {Idomura},\ and\ \citenamefont {Villard}}]{JOLLIET2007}%
  \BibitemOpen
  \bibfield  {author} {\bibinfo {author} {\bibfnamefont {S.}~\bibnamefont {Jolliet}}, \bibinfo {author} {\bibfnamefont {A.}~\bibnamefont {Bottino}}, \bibinfo {author} {\bibfnamefont {P.}~\bibnamefont {Angelino}}, \bibinfo {author} {\bibfnamefont {R.}~\bibnamefont {Hatzky}}, \bibinfo {author} {\bibfnamefont {T.~M.}\ \bibnamefont {Tran}}, \bibinfo {author} {\bibfnamefont {B.~F.}\ \bibnamefont {Mcmillan}}, \bibinfo {author} {\bibfnamefont {O.}~\bibnamefont {Sauter}}, \bibinfo {author} {\bibfnamefont {K.}~\bibnamefont {Appert}}, \bibinfo {author} {\bibfnamefont {Y.}~\bibnamefont {Idomura}}, \ and\ \bibinfo {author} {\bibfnamefont {L.}~\bibnamefont {Villard}},\ }\bibfield  {title} {\enquote {\bibinfo {title} {A global collisionless pic code in magnetic coordinates},}\ }\href@noop {} {\bibfield  {journal} {\bibinfo  {journal} {Comput. Phys. Comm.}\ }\textbf {\bibinfo {volume} {177}},\ \bibinfo {pages} {409} (\bibinfo {year} {2007})}\BibitemShut {NoStop}%
\bibitem [{\citenamefont {Idomura}\ \emph {et~al.}(2008)\citenamefont {Idomura}, \citenamefont {Ida}, \citenamefont {Kano}, \citenamefont {Aiba},\ and\ \citenamefont {Tokuda}}]{IDOMURA2008}%
  \BibitemOpen
  \bibfield  {author} {\bibinfo {author} {\bibfnamefont {Y.}~\bibnamefont {Idomura}}, \bibinfo {author} {\bibfnamefont {M.}~\bibnamefont {Ida}}, \bibinfo {author} {\bibfnamefont {T.}~\bibnamefont {Kano}}, \bibinfo {author} {\bibfnamefont {N.}~\bibnamefont {Aiba}}, \ and\ \bibinfo {author} {\bibfnamefont {S.}~\bibnamefont {Tokuda}},\ }\bibfield  {title} {\enquote {\bibinfo {title} {Conservative global gyrokinetic toroidal full-f five-dimensional vlasov simulation},}\ }\href@noop {} {\bibfield  {journal} {\bibinfo  {journal} {Comput. Phys. Comm.}\ }\textbf {\bibinfo {volume} {179}},\ \bibinfo {pages} {391} (\bibinfo {year} {2008})}\BibitemShut {NoStop}%
\bibitem [{\citenamefont {Obrejan}\ \emph {et~al.}(2015)\citenamefont {Obrejan}, \citenamefont {Imadera}, \citenamefont {Li},\ and\ \citenamefont {Kishimoto}}]{OBREJAN2015}%
  \BibitemOpen
  \bibfield  {author} {\bibinfo {author} {\bibfnamefont {K.}~\bibnamefont {Obrejan}}, \bibinfo {author} {\bibfnamefont {K.}~\bibnamefont {Imadera}}, \bibinfo {author} {\bibfnamefont {J.}~\bibnamefont {Li}}, \ and\ \bibinfo {author} {\bibfnamefont {Y.}~\bibnamefont {Kishimoto}},\ }\bibfield  {title} {\enquote {\bibinfo {title} {Development of a global toroidal gyrokinetic vlasov code with new real space field solver},}\ }\href@noop {} {\bibfield  {journal} {\bibinfo  {journal} {Plasma Fusion Res.}\ }\textbf {\bibinfo {volume} {10}},\ \bibinfo {pages} {3403042} (\bibinfo {year} {2015})}\BibitemShut {NoStop}%
\bibitem [{\citenamefont {Dai}\ \emph {et~al.}(2019)\citenamefont {Dai}, \citenamefont {Xu}, \citenamefont {Ye}, \citenamefont {Xiao},\ and\ \citenamefont {Wang}}]{DAI2019}%
  \BibitemOpen
  \bibfield  {author} {\bibinfo {author} {\bibfnamefont {Z.}~\bibnamefont {Dai}}, \bibinfo {author} {\bibfnamefont {Y.}~\bibnamefont {Xu}}, \bibinfo {author} {\bibfnamefont {L.}~\bibnamefont {Ye}}, \bibinfo {author} {\bibfnamefont {X.}~\bibnamefont {Xiao}}, \ and\ \bibinfo {author} {\bibfnamefont {S.}~\bibnamefont {Wang}},\ }\bibfield  {title} {\enquote {\bibinfo {title} {Gyrokinetic simulation of itg turbulence with toroidal geometry including the magnetic axis by using field-aligned coordinates},}\ }\href@noop {} {\bibfield  {journal} {\bibinfo  {journal} {Comput. Phys. Comm.}\ }\textbf {\bibinfo {volume} {242}},\ \bibinfo {pages} {72} (\bibinfo {year} {2019})}\BibitemShut {NoStop}%
\bibitem [{\citenamefont {Matsuoka}, \citenamefont {Idomura},\ and\ \citenamefont {Satake}(2018)}]{Matsuoka2018POP}%
  \BibitemOpen
  \bibfield  {author} {\bibinfo {author} {\bibfnamefont {S.}~\bibnamefont {Matsuoka}}, \bibinfo {author} {\bibfnamefont {S.}~\bibnamefont {Idomura}}, \ and\ \bibinfo {author} {\bibfnamefont {S.}~\bibnamefont {Satake}},\ }\bibfield  {title} {\enquote {\bibinfo {title} {{Neoclassical transport benchmark of global full-f gyrokinetic simulation in stellarator configurations}},}\ }\href@noop {} {\bibfield  {journal} {\bibinfo  {journal} {Phys. Plasmas}\ }\textbf {\bibinfo {volume} {25}},\ \bibinfo {pages} {022510} (\bibinfo {year} {2018})}\BibitemShut {NoStop}%
\bibitem [{\citenamefont {Feng}\ \emph {et~al.}(2018)\citenamefont {Feng}, \citenamefont {Zhang}, \citenamefont {Lin}, \citenamefont {Zhufu}, \citenamefont {Xu}, \citenamefont {Cao},\ and\ \citenamefont {Li}}]{FENG2018}%
  \BibitemOpen
  \bibfield  {author} {\bibinfo {author} {\bibfnamefont {H.}~\bibnamefont {Feng}}, \bibinfo {author} {\bibfnamefont {W.}~\bibnamefont {Zhang}}, \bibinfo {author} {\bibfnamefont {Z.}~\bibnamefont {Lin}}, \bibinfo {author} {\bibfnamefont {X.}~\bibnamefont {Zhufu}}, \bibinfo {author} {\bibfnamefont {J.}~\bibnamefont {Xu}}, \bibinfo {author} {\bibfnamefont {J.}~\bibnamefont {Cao}}, \ and\ \bibinfo {author} {\bibfnamefont {D.}~\bibnamefont {Li}},\ }\bibfield  {title} {\enquote {\bibinfo {title} {Development of finite element field solver in gyrokinetic toroidal code},}\ }\href@noop {} {\bibfield  {journal} {\bibinfo  {journal} {Commun. Comput. Phys.}\ }\textbf {\bibinfo {volume} {24}},\ \bibinfo {pages} {655} (\bibinfo {year} {2018})}\BibitemShut {NoStop}%
\bibitem [{\citenamefont {Ye}\ \emph {et~al.}(2016)\citenamefont {Ye}, \citenamefont {Xu}, \citenamefont {Xiao}, \citenamefont {Dai},\ and\ \citenamefont {Wang}}]{Ye2016}%
  \BibitemOpen
  \bibfield  {author} {\bibinfo {author} {\bibfnamefont {L.}~\bibnamefont {Ye}}, \bibinfo {author} {\bibfnamefont {Y.}~\bibnamefont {Xu}}, \bibinfo {author} {\bibfnamefont {X.}~\bibnamefont {Xiao}}, \bibinfo {author} {\bibfnamefont {Z.}~\bibnamefont {Dai}}, \ and\ \bibinfo {author} {\bibfnamefont {S.}~\bibnamefont {Wang}},\ }\bibfield  {title} {\enquote {\bibinfo {title} {A gyrokinetic continuum code based on the numerical lie transform ({NLT}) method},}\ }\href@noop {} {\bibfield  {journal} {\bibinfo  {journal} {J. Comput. Phys.}\ }\textbf {\bibinfo {volume} {316}},\ \bibinfo {pages} {180} (\bibinfo {year} {2016})}\BibitemShut {NoStop}%
\bibitem [{\citenamefont {Xu}\ \emph {et~al.}(2017)\citenamefont {Xu}, \citenamefont {Ye}, \citenamefont {Dai}, \citenamefont {Xiao},\ and\ \citenamefont {Wang}}]{XU2017}%
  \BibitemOpen
  \bibfield  {author} {\bibinfo {author} {\bibfnamefont {Y.}~\bibnamefont {Xu}}, \bibinfo {author} {\bibfnamefont {L.}~\bibnamefont {Ye}}, \bibinfo {author} {\bibfnamefont {Z.}~\bibnamefont {Dai}}, \bibinfo {author} {\bibfnamefont {X.}~\bibnamefont {Xiao}}, \ and\ \bibinfo {author} {\bibfnamefont {S.}~\bibnamefont {Wang}},\ }\bibfield  {title} {\enquote {\bibinfo {title} {{Nonlinear gyrokinetic simulation of ion temperature gradient turbulence based on a numerical Lie-transform perturbation method}},}\ }\href@noop {} {\bibfield  {journal} {\bibinfo  {journal} {Phys. Plasmas}\ }\textbf {\bibinfo {volume} {24}},\ \bibinfo {pages} {082515} (\bibinfo {year} {2017})}\BibitemShut {NoStop}%
\bibitem [{\citenamefont {Wang}(2012)}]{WANG2012}%
  \BibitemOpen
  \bibfield  {author} {\bibinfo {author} {\bibfnamefont {S.}~\bibnamefont {Wang}},\ }\bibfield  {title} {\enquote {\bibinfo {title} {{Transport formulation of the gyrokinetic turbulence}},}\ }\href@noop {} {\bibfield  {journal} {\bibinfo  {journal} {Phys. Plasmas}\ }\textbf {\bibinfo {volume} {19}},\ \bibinfo {pages} {062504} (\bibinfo {year} {2012})}\BibitemShut {NoStop}%
\bibitem [{\citenamefont {Wang}(2013{\natexlab{a}})}]{WANG2013POP}%
  \BibitemOpen
  \bibfield  {author} {\bibinfo {author} {\bibfnamefont {S.}~\bibnamefont {Wang}},\ }\bibfield  {title} {\enquote {\bibinfo {title} {{Nonlinear scattering term in the gyrokinetic Vlasov equation}},}\ }\href@noop {} {\bibfield  {journal} {\bibinfo  {journal} {Phys. Plasmas}\ }\textbf {\bibinfo {volume} {20}},\ \bibinfo {pages} {082312} (\bibinfo {year} {2013}{\natexlab{a}})}\BibitemShut {NoStop}%
\bibitem [{\citenamefont {Wang}(2013{\natexlab{b}})}]{Wang2013PRE}%
  \BibitemOpen
  \bibfield  {author} {\bibinfo {author} {\bibfnamefont {S.}~\bibnamefont {Wang}},\ }\bibfield  {title} {\enquote {\bibinfo {title} {{Kinetic theory of weak turbulence in plasmas}},}\ }\href@noop {} {\bibfield  {journal} {\bibinfo  {journal} {Phys. Rev. E.}\ }\textbf {\bibinfo {volume} {87}},\ \bibinfo {pages} {063103} (\bibinfo {year} {2013}{\natexlab{b}})}\BibitemShut {NoStop}%
\bibitem [{\citenamefont {Xu}, \citenamefont {Dai},\ and\ \citenamefont {Wang}(2014)}]{Xu2014POP}%
  \BibitemOpen
  \bibfield  {author} {\bibinfo {author} {\bibfnamefont {Y.}~\bibnamefont {Xu}}, \bibinfo {author} {\bibfnamefont {Z.}~\bibnamefont {Dai}}, \ and\ \bibinfo {author} {\bibfnamefont {S.}~\bibnamefont {Wang}},\ }\bibfield  {title} {\enquote {\bibinfo {title} {{Nonlinear gyrokinetic theory based on a new method and computation of the guiding-center orbit in tokamaks}},}\ }\href@noop {} {\bibfield  {journal} {\bibinfo  {journal} {Phys. Plasmas}\ }\textbf {\bibinfo {volume} {21}},\ \bibinfo {pages} {042505} (\bibinfo {year} {2014})}\BibitemShut {NoStop}%
\bibitem [{\citenamefont {Lewis}\ and\ \citenamefont {Bellan}(1990)}]{LEWIS1990}%
  \BibitemOpen
  \bibfield  {author} {\bibinfo {author} {\bibfnamefont {H.~R.}\ \bibnamefont {Lewis}}\ and\ \bibinfo {author} {\bibfnamefont {P.~M.}\ \bibnamefont {Bellan}},\ }\bibfield  {title} {\enquote {\bibinfo {title} {{Physical constraints on the coefficients of Fourier expansions in cylindrical coordinates}},}\ }\href@noop {} {\bibfield  {journal} {\bibinfo  {journal} {J. Math. Phys.}\ }\textbf {\bibinfo {volume} {31}},\ \bibinfo {pages} {2592} (\bibinfo {year} {1990})}\BibitemShut {NoStop}%
\bibitem [{\citenamefont {Eisen}, \citenamefont {Heinrichs},\ and\ \citenamefont {Witsch}(1991)}]{EISEN1991}%
  \BibitemOpen
  \bibfield  {author} {\bibinfo {author} {\bibfnamefont {H.}~\bibnamefont {Eisen}}, \bibinfo {author} {\bibfnamefont {W.}~\bibnamefont {Heinrichs}}, \ and\ \bibinfo {author} {\bibfnamefont {K.}~\bibnamefont {Witsch}},\ }\bibfield  {title} {\enquote {\bibinfo {title} {Spectral collocation methods and polar coordinate singularities},}\ }\href@noop {} {\bibfield  {journal} {\bibinfo  {journal} {J. Comput. Phys.}\ }\textbf {\bibinfo {volume} {96}},\ \bibinfo {pages} {241} (\bibinfo {year} {1991})}\BibitemShut {NoStop}%
\bibitem [{\citenamefont {Xiao}\ \emph {et~al.}(2017)\citenamefont {Xiao}, \citenamefont {Ye}, \citenamefont {Xu},\ and\ \citenamefont {Wang}}]{Xiao2017CCP}%
  \BibitemOpen
  \bibfield  {author} {\bibinfo {author} {\bibfnamefont {X.}~\bibnamefont {Xiao}}, \bibinfo {author} {\bibfnamefont {L.}~\bibnamefont {Ye}}, \bibinfo {author} {\bibfnamefont {Y.}~\bibnamefont {Xu}}, \ and\ \bibinfo {author} {\bibfnamefont {S.}~\bibnamefont {Wang}},\ }\bibfield  {title} {\enquote {\bibinfo {title} {{Application of High Dimensional B-Spline Interpolation in Solving the Gyro-Kinetic Vlasov Equation Based on Semi-Lagrangian Method}},}\ }\href@noop {} {\bibfield  {journal} {\bibinfo  {journal} {Commun. Comput. Phys.}\ }\textbf {\bibinfo {volume} {22}},\ \bibinfo {pages} {789} (\bibinfo {year} {2017})}\BibitemShut {NoStop}%
\bibitem [{\citenamefont {Hazeltine}\ and\ \citenamefont {Meiss}(1992)}]{Hazeltine1988}%
  \BibitemOpen
  \bibfield  {author} {\bibinfo {author} {\bibfnamefont {R.~D.}\ \bibnamefont {Hazeltine}}\ and\ \bibinfo {author} {\bibfnamefont {J.~D.}\ \bibnamefont {Meiss}},\ }\href@noop {} {\emph {\bibinfo {title} {Plasma Confinement}}}\ (\bibinfo  {publisher} {Addison-Wesley Publishing Company},\ \bibinfo {year} {1992})\ pp.\ \bibinfo {pages} {52--55}\BibitemShut {NoStop}%
\bibitem [{\citenamefont {Wang}(2017)}]{WANG2017}%
  \BibitemOpen
  \bibfield  {author} {\bibinfo {author} {\bibfnamefont {S.}~\bibnamefont {Wang}},\ }\bibfield  {title} {\enquote {\bibinfo {title} {{Zonal flows driven by the turbulent energy flux and the turbulent toroidal Reynolds stress in a magnetic fusion torus}},}\ }\href@noop {} {\bibfield  {journal} {\bibinfo  {journal} {Phys. Plasmas}\ }\textbf {\bibinfo {volume} {24}},\ \bibinfo {pages} {102508} (\bibinfo {year} {2017})}\BibitemShut {NoStop}%
\bibitem [{\citenamefont {Burrell}\ \emph {et~al.}(1998)\citenamefont {Burrell}, \citenamefont {Austin}, \citenamefont {Greenfield}, \citenamefont {Lao}, \citenamefont {Rice}, \citenamefont {Staebler},\ and\ \citenamefont {Stallard}}]{BURREL1998}%
  \BibitemOpen
  \bibfield  {author} {\bibinfo {author} {\bibfnamefont {K.~H.}\ \bibnamefont {Burrell}}, \bibinfo {author} {\bibfnamefont {M.~E.}\ \bibnamefont {Austin}}, \bibinfo {author} {\bibfnamefont {C.~M.}\ \bibnamefont {Greenfield}}, \bibinfo {author} {\bibfnamefont {L.~L.}\ \bibnamefont {Lao}}, \bibinfo {author} {\bibfnamefont {B.~W.}\ \bibnamefont {Rice}}, \bibinfo {author} {\bibfnamefont {G.~M.}\ \bibnamefont {Staebler}}, \ and\ \bibinfo {author} {\bibfnamefont {B.~W.}\ \bibnamefont {Stallard}},\ }\bibfield  {title} {\enquote {\bibinfo {title} {Effects of velocity shear and magnetic shear in the formation of core transport barriers in the diii-d tokamak},}\ }\href@noop {} {\bibfield  {journal} {\bibinfo  {journal} {Plasma Phys. Control. Fusion}\ }\textbf {\bibinfo {volume} {40}},\ \bibinfo {pages} {1585} (\bibinfo {year} {1998})}\BibitemShut {NoStop}%
\bibitem [{\citenamefont {Lele}(1992)}]{LELE1992}%
  \BibitemOpen
  \bibfield  {author} {\bibinfo {author} {\bibfnamefont {S.~K.}\ \bibnamefont {Lele}},\ }\bibfield  {title} {\enquote {\bibinfo {title} {Compact finite difference schemes with spectral-like resolution},}\ }\href@noop {} {\bibfield  {journal} {\bibinfo  {journal} {J. Comput. Phys.}\ }\textbf {\bibinfo {volume} {103}},\ \bibinfo {pages} {16} (\bibinfo {year} {1992})}\BibitemShut {NoStop}%
\bibitem [{\citenamefont {Zonca}\ and\ \citenamefont {Chen}(2008)}]{ZONCA2008}%
  \BibitemOpen
  \bibfield  {author} {\bibinfo {author} {\bibfnamefont {F.}~\bibnamefont {Zonca}}\ and\ \bibinfo {author} {\bibfnamefont {L.}~\bibnamefont {Chen}},\ }\bibfield  {title} {\enquote {\bibinfo {title} {Radial structures and nonlinear excitation of geodesic acoustic modes},}\ }\href@noop {} {\bibfield  {journal} {\bibinfo  {journal} {Euro. Phys. Lett.}\ }\textbf {\bibinfo {volume} {83}},\ \bibinfo {pages} {35001} (\bibinfo {year} {2008})}\BibitemShut {NoStop}%
\bibitem [{\citenamefont {Rosenbluth}\ and\ \citenamefont {Hinton}(1998)}]{ROUSENBLUTH1998}%
  \BibitemOpen
  \bibfield  {author} {\bibinfo {author} {\bibfnamefont {M.~N.}\ \bibnamefont {Rosenbluth}}\ and\ \bibinfo {author} {\bibfnamefont {F.~L.}\ \bibnamefont {Hinton}},\ }\bibfield  {title} {\enquote {\bibinfo {title} {Poloidal flow driven by ion-temperature-gradient turbulence in tokamaks},}\ }\href@noop {} {\bibfield  {journal} {\bibinfo  {journal} {Phys. Rev. Lett.}\ }\textbf {\bibinfo {volume} {80}},\ \bibinfo {pages} {724} (\bibinfo {year} {1998})}\BibitemShut {NoStop}%
\bibitem [{\citenamefont {Sugama}\ and\ \citenamefont {Watanabe}(2006)}]{SUGAMA2006}%
  \BibitemOpen
  \bibfield  {author} {\bibinfo {author} {\bibfnamefont {H.}~\bibnamefont {Sugama}}\ and\ \bibinfo {author} {\bibfnamefont {T.-H.}\ \bibnamefont {Watanabe}},\ }\bibfield  {title} {\enquote {\bibinfo {title} {Collisionless damping of geodesic acoustic modes},}\ }\href@noop {} {\bibfield  {journal} {\bibinfo  {journal} {J. Plasma Phys.}\ }\textbf {\bibinfo {volume} {72}},\ \bibinfo {pages} {825} (\bibinfo {year} {2006})}\BibitemShut {NoStop}%
\bibitem [{\citenamefont {Sugama}\ and\ \citenamefont {Watanabe}(2008)}]{Sugama2008}%
  \BibitemOpen
  \bibfield  {author} {\bibinfo {author} {\bibfnamefont {H.}~\bibnamefont {Sugama}}\ and\ \bibinfo {author} {\bibfnamefont {T.-H.}\ \bibnamefont {Watanabe}},\ }\bibfield  {title} {\enquote {\bibinfo {title} {Erratum: ‘collisionless damping of geodesic acoustic modes’ [j. plasma physics (2006) 72, 825]},}\ }\href@noop {} {\bibfield  {journal} {\bibinfo  {journal} {J. Plasma Phys.}\ }\textbf {\bibinfo {volume} {74}},\ \bibinfo {pages} {139} (\bibinfo {year} {2008})}\BibitemShut {NoStop}%
\bibitem [{\citenamefont {Dimits}\ \emph {et~al.}(2000)\citenamefont {Dimits}, \citenamefont {Bateman}, \citenamefont {Beer}, \citenamefont {Cohen}, \citenamefont {Dorland}, \citenamefont {Hammett}, \citenamefont {Kim}, \citenamefont {Kinsey}, \citenamefont {Kotschenreuther}, \citenamefont {Kritz}, \citenamefont {Lao}, \citenamefont {Mandrekas}, \citenamefont {Nevins}, \citenamefont {Parker}, \citenamefont {Redd}, \citenamefont {Shumaker}, \citenamefont {Sydora},\ and\ \citenamefont {Weiland}}]{Dimits2000}%
  \BibitemOpen
  \bibfield  {author} {\bibinfo {author} {\bibfnamefont {A.~M.}\ \bibnamefont {Dimits}}, \bibinfo {author} {\bibfnamefont {G.}~\bibnamefont {Bateman}}, \bibinfo {author} {\bibfnamefont {M.~A.}\ \bibnamefont {Beer}}, \bibinfo {author} {\bibfnamefont {B.~I.}\ \bibnamefont {Cohen}}, \bibinfo {author} {\bibfnamefont {W.}~\bibnamefont {Dorland}}, \bibinfo {author} {\bibfnamefont {G.~W.}\ \bibnamefont {Hammett}}, \bibinfo {author} {\bibfnamefont {C.}~\bibnamefont {Kim}}, \bibinfo {author} {\bibfnamefont {J.~E.}\ \bibnamefont {Kinsey}}, \bibinfo {author} {\bibfnamefont {M.}~\bibnamefont {Kotschenreuther}}, \bibinfo {author} {\bibfnamefont {A.~H.}\ \bibnamefont {Kritz}}, \bibinfo {author} {\bibfnamefont {L.~L.}\ \bibnamefont {Lao}}, \bibinfo {author} {\bibfnamefont {J.}~\bibnamefont {Mandrekas}}, \bibinfo {author} {\bibfnamefont {W.~M.}\ \bibnamefont {Nevins}}, \bibinfo {author} {\bibfnamefont {S.~E.}\ \bibnamefont {Parker}}, \bibinfo {author} {\bibfnamefont {A.~J.}\ \bibnamefont {Redd}}, \bibinfo {author}
  {\bibfnamefont {D.~E.}\ \bibnamefont {Shumaker}}, \bibinfo {author} {\bibfnamefont {R.}~\bibnamefont {Sydora}}, \ and\ \bibinfo {author} {\bibfnamefont {J.}~\bibnamefont {Weiland}},\ }\bibfield  {title} {\enquote {\bibinfo {title} {{Comparisons and physics basis of tokamak transport models and turbulence simulations}},}\ }\href@noop {} {\bibfield  {journal} {\bibinfo  {journal} {Phys. Plasmas}\ }\textbf {\bibinfo {volume} {7}},\ \bibinfo {pages} {969} (\bibinfo {year} {2000})}\BibitemShut {NoStop}%
\end{thebibliography}%

\end{document}